\documentclass[aps,showpacs]{revtex4}
\usepackage{graphicx}
\usepackage{epsfig}
\usepackage{amsmath}
\usepackage{graphics}
\usepackage{latexsym}
\usepackage{subfigure} 

\begin{document} 

\title{Semiflexible Polymers Under Good Solvent Conditions\\ Interacting With 
Repulsive Walls}
\author{Sergei A. Egorov$^{1,3,4}$, Andrey Milchev$^2$, Peter Virnau$^3$, and 
Kurt Binder$^3$}

\affiliation{$^1$ Department of Chemistry, University of Virginia,
Charlottesville, VA 22901, USA [E-Mail:sae6z$@$cms.mail.virginia.edu] \\
$^2$ Institute for Physical Chemistry, Bulgarian Academia of Sciences, 1113
Sofia, Bulgaria \\
$^3$ Institut f\"ur Physik, Johannes Gutenberg Universit\"at Mainz, 55099 Mainz, 
Germany\\ $^4$ Leibniz-Institut f\"ur Polymerforschung, Institut Theorie der 
Polymere,\\ Hohe Str. 6, 01069 Dresden, Germany}

\begin{abstract}
Solutions of semiflexible polymers confined by repulsive planar walls are 
studied by density functional theory and Molecular Dynamics simulations, to 
clarify the competition between the chain alignment favored by the wall and the 
depletion caused by the monomer-wall repulsion. A coarse-grained bead-spring 
model with bond bending potential is studied, varying both the contour length 
and the persistence length of the polymers, as well as the monomer concentration 
in the solution (good solvent conditions are assumed throughout, and solvent 
molecules are not included explicitly). The profiles of monomer density and 
pressure tensor components near the wall are studied, and the surface tension of 
the solution is obtained. While the surface tension slightly decreases with 
chain length for flexible polymers, it clearly increases with chain length for 
stiff polymers. Thus, at fixed density and fixed chain length the surface 
tension also increases with increasing persistence length. Chain ends always are 
enriched near the wall, but this effect is much larger for stiff polymers than 
for flexible ones. Also the profiles of the mean square gyration radius 
components near the wall and the nematic order parameter are studied to clarify 
the conditions where wall-induced nematic order occurs.
\end{abstract}

\maketitle

\section{Introduction}
\label{section1}

The interplay of stiffness on the local scale and flexibility on larger scales
along the contour of macromolecules is a crucial aspect to understand their
conformations in solutions and melts and their resulting physical
properties~\cite{flory69,degennes79,grosberg94}. The intrinsic stiffness of a
polymer is normally characterized by its persistence
length~\cite{flory69,degennes79,grosberg94,hsu10} $\ell _p$, and while for
flexible polymers $\ell_p$ is of the same order as the length $\ell _b$ of the
effective bond between subsequent (effective) monomeric units, for semiflexible
polymers one has $\ell_p \gg \ell_b$, and such semiflexible polymers are
particularly common in a biophysical context. E.g., for double stranded (ds) DNA
in typical cases~\cite{bustamante94} $\ell_p \approx 50 nm$ and~\cite{parry84}
$\ell_b \approx 0.26 - 0.5 nm$, and objects such as actin filaments are even
much stiffer. Hence there is interest to study the full regime of contour
lengths $L=(N-1)\ell_b$, $N$ being the number of effective monomers which we
henceforth shall refer to as ``chain length'', from $L \gg \ell_p$, where the
chain in dilute solution behaves like a random coil~\cite{degennes79}, to the
inverse limit $\ell _p \gg L$, where the polymer behaves almost like  a rigid
rod. As is well known, thin long rigid rods in solution under good solvent
conditions undergo an entropically driven transition from an isotropic phase to
nematic order~\cite{onsager49,degennes92}. However, also in the case when $L$ is
of the same order as $\ell_p$ or larger, liquid crystalline order can occur in
solutions of semiflexible polymers, and this problem has found longstanding
attention by theory~\cite{khokhlov81,khokhlov82,semenov88,wilson93} as well as
experimentally~\cite{ciferri83} and is relevant for many applications, e.g.
displays, fibers with high mechanical strength, microelectromechanical and
biomedical devices~\cite{grell97,wang09b,donald06,woltman07}. In many
circumstances, the interaction of the semiflexible polymers with confining walls
is a crucial
aspect~\cite{yethiraj94,chen95b,escobedo97,micheletti05,chen05c,chen07,
turesson07,ivanov13,ivanov14} but clearly this problem is not yet fully 
understood (even predicting nematic phases of semiflexible polymers in the bulk
still is under current study~\cite{zhang15}).

In the present paper, we take a step towards the better understanding of
solutions of semiflexible polymers interacting with repulsive walls, focusing on
the case where in the bulk the solution is always in the isotropic fluid phase.
We shall study a coarse-grained off-lattice model that shall be described in
Sec.~II, together with a brief characterization of our methods (Molecular
Dynamics (MD)  simulations~\cite{allen89,rapaport04} and density functional
theory in a formulation appropriate for semiflexible macromolecules). A study of
the isotropic-nematic phase transition of this model in the bulk will be
presented elsewhere.~\cite{egorov16} Sec.~III gives typical MD results, while
Sec. IV presents our DFT results, where extensive variation of the persistence
length, the chain length and the polymer concentration in the bulk solution will
be given. Due to excessive demands in computer resources it would be premature
to attempt such a study by MD methods alone; however, the direct comparison of
MD and DFT results for the same coarse-grained model serves to ascertain the
accuracy of the DFT results and to clarify their limitations. Sec. V gives a
discussion of our results and an outlook on open problems.

\section{Model, Methods and Some Theoretical Background}
\label{section2}
\subsection{Semiflexible Polymers: A coarse-grained model, and pertinent 
theoretical results} \label{subsection2a}
The standard coarse-grained off-lattice model for flexible macromolecules is the 
bead-spring model where subsequent monomers along the chain interact both with 
the ``finite extensible nonlinear elastic'' (FENE) 
potential~\cite{grest86,kremer90}
\begin{equation}
V^{FENE}(r) = -0.5 kr_0^2 \ln [1-(r/r_0)^2], \; r <r_0,
\label{eq1}
\end{equation}

$V^{FENE}(r>r_0)\equiv 0$, and a purely repulsive 
Weeks-Chandler-Andersen~\cite{weeks71} (WCA)-type potential,

\begin{equation}
V^{WCA}(r) = 4 \epsilon [(\sigma /r)^{12}-(\sigma/r)^6+1/4], \; r<r_c = \sigma
2^{1/6},
\label{eq2}
\end{equation}

and $V^{WCA}(r>r_c)=0$. Eq.~\ref{eq2} also acts between any pair of beads, that
represent the effective monomeric units of a chain, in the system. This purely
repulsive interaction between any pair of monomers represents the dominance of
excluded volume interactions in a polymer solution under very good solvent
conditions in the framework of this ``implicit solvent'' model, where solvent
molecules are not considered explicitly. If one plots the combined Kremer-Grest potential, that is, the sum of Eqs.~\ref{eq1} 
and ~\ref{eq2}, that defines the bond length in the bead-spring model, one finds a
minimum of the potential well at $\approx 0.96$, which is then confirmed by
analyzing simulation data. Since the well is itself rather steep, this
effective length of the bonds is rather insensitive to parameters (like
concentration, or temperature) variation. 

We choose units of length $\sigma = 1$ and energy $\epsilon = 1$ and employ a
temperature $k_BT=1$ as well. The parameters $k$ and $r_0$ then are chosen as
$r_0=1.5 \sigma $ and $k=30 \epsilon/\sigma^2$=30. The distance between beads
along the chain then is $\ell_b \approx 0.96.$ As a potential due to the repulsive walls, we use a potential of the same functional form as Eq.~(\ref{eq2}), 
$V^{WCA}(z)$, where $z$ is the distance to the closest wall. 

In order to consider semiflexible rather than fully flexible polymers, we
augment  Eqs.~\ref{eq1},\ref{eq2} by a bond bending potential,
\begin{equation}
V_{bend} (\theta_{ijk}) = \epsilon_b[1-\cos(\theta_{ijk})],
\label{eq3}
\end{equation}

where $\theta_{ijk}$ is the bond angle formed between the two subsequent unit
vectors along the bonds connecting monomers $i,j=i+1$ and $j,k =i+2$. The energy
parameter $\epsilon_b$ then controls the persistence length $\ell_p$, which is
defined here as~\cite{hsu10}
\begin{equation}
\ell_p/\ell_b = -1/\ln \langle \cos \theta _{ijk}\rangle
\label{eq4}
\end{equation}

We recall that for a ``phantom chain'' (i.e. excluded volume interactions being
strictly zero) Eq.~\ref{eq4} is fully equivalent to the traditional textbook
definition~\cite{grosberg94} of $\ell_p$ as a decay constant of bond
orientational correlations along the chain contour, $\vec{a}_i$ being the bond
vector connecting monomers i and $i+1$,
\begin{equation}
\langle \vec{a}_i \cdot \vec{a}_{i+s}\rangle = \langle \vec{a}_i^2\rangle \exp (- \ell_bs/\ell_p).
\label{eq5}
\end{equation}

In our case, Eq.~\ref{eq5} is not useful since in dilute solution under good
solvent conditions for large $s$ we rather have a power-law
decay~\cite{grosberg94,schaefer99}

\begin{equation}
\langle \vec{a}_i \cdot \vec{a}_{i+s}\rangle \propto s^{- \beta}\quad N^* \ll s
\ll N,
\label{eq6}
\end{equation}

where $\beta = 2-2\nu$, $\nu\approx 0.588$ being the exponent characterizing the
mean-square end-to-end distance of a coil~\cite{leguillou80}, $\langle
R_e^2\rangle \propto N ^{2 \nu}$. The chain length $N^*$ that characterizes the 
onset of excluded volume effects for semiflexible chains is~\cite{hsu12}, in 
$d=3$ dimensions, $N^* \approx \ell_p^3/(\ell_bd^2_m)=(\ell_p/\ell_b)^3$, 
$d_m\approx\ell_b$ being the diameter of an effective monomer. While for 
flexible polymers ($\epsilon_b/k_BT=0$ in Eq.~\ref{eq3}) we have $\ell_p \approx 
\ell_b$ and hence excluded volume dominates already for short chains, for 
$\epsilon_b/k_BT \geq 2$ we have $\ell_p /\ell_b \approx \epsilon_b/k_BT$, and 
thus for large $\epsilon_b/k_BT$ and moderate chain lengths we can reach 
conditions where excluded volume effects do not matter much. Only when we 
encounter wall-attached chains (i.e., $d=2$ dimensions), we would have $\nu = 
3/4$ and $N^* \approx \ell_p/\ell_b$, i.e. excluded volume effects set in 
already at the crossover from rods to coils. However, such wall-attached 
polymers for repulsive monomer-wall interactions are not expected for dilute 
solutions, but might occur only for higher polymer 
concentrations~\cite{ivanov13,ivanov14}. We note that for concentrated solutions 
and melts we have~\cite{shirvanyants08,wittmer04} $\beta = 3/2$ in 
Eq.~\ref{eq6}, and~\cite{kratky49} $\langle R_e^2\rangle \approx 2 
\ell_b\ell_pN$ for $N \gg \ell_p/\ell_b$. We also recall that for conditions for 
which excluded volume interactions are not dominant the Kratky-Porod wormlike 
chain model~\cite{kratky49} is expected to describe correctly the crossover from 
rod-like chains $(\langle R_e^2\rangle \approx \ell_b^2N^2)$ to Gaussian coils,

\begin{equation}
\langle R_e^2\rangle = 2 \ell_p L\{1-\frac {\ell_p}{L}
[1-\exp(-L/\ell_p)]\}\quad , \quad L=(N-1)\ell_b\quad .
\label{eq7}
\end{equation}

Thus, when we vary chain length $N$, persistence length $\ell_p$ and
concentration of the polymer solution, we must be aware that many distinct
regimes with different types of behavior may play a role. Of particular
interest, of course, is the regime where $L$ and $\ell_p$ are comparable, and
for high enough concentration onset of nematic order is expected in the bulk. 
For the two limiting cases ($L \ll \ell_p$ and $L \gg \ell_p$), the nematic order starts when the
monomer concentration $\rho$ exceeds a critical value $\rho_i$~\cite{onsager49,khokhlov81,khokhlov82,semenov88}
\begin{eqnarray}
\rho_i\ell_p/d\propto \begin{cases} const, &L \gg \ell_p\\ 
                              \ell_p/L, &L \ll \ell_p\end{cases}
\label{eq8}
\end{eqnarray}
In the present work only concentration $\rho < \rho_{i}$ will be studied, however.

Of course, the model defined by Eqs.~\ref{eq1} - \ref{eq3} is not the only
coarse-grained model of a semi-flexible polymer that is conceivable. Another
useful model (studied e.g. by Yethiraj~\cite{yethiraj94}) considers a chain of
tangent hard spheres, where stiffness again is controlled by the potential,
Eq.~\ref{eq3}. While this model is very useful in a  DFT context,  it is less
convenient for MD simulation.

\subsection{Some Technical Aspects of the MD simulations}
\label{subsection2b}
MD simulations were carried out both on single CPU's using a code written by
us~\cite{egorov15a,milchev15}, as well as on graphics processing units (GPU's),
using the HooMD Blue software~\cite{anderson08,glaser15}. For large enough
systems, a speedup of about a factor of 100 was gained by the use of GPU's.

In the MD simulations the Newton equations of motion of the many-particle system are integrated numerically, applying the standard Velocity
Verlet algorithm~\cite{allen89,rapaport04}. In order to work in the constant temperature rather than constant energy ensemble, a Langevin thermostat is used, i.e.
the positions $\vec{r}_n(t)$ of the effective monomers evolve with time
according to~\cite{grest86}
\begin{equation}
m \frac {d^2\vec{r}_n}{dt^2} = \vec{F}_{tot} (\vec{r}_n) - \gamma \frac
{d\vec{r}_n}{dt} + \vec{F}_n^{rand}(t)\;,
\label{eq9}
\end{equation}

where the mass $m=1$ of a monomeric unit leads to a unit of time $\tau _{MD}=
\sqrt{m\sigma^2/\epsilon} =1$ as well. The friction coefficient $\gamma =
0.25$ is chosen, and sets the scale for the random force $\vec{F}_n^{rand}(t)$
via the fluctuation-dissipation relation as usual,
\begin{equation}
\langle \vec{F}_n^{rand} (t) \cdot \vec{F}_{n'}^{rand} (t')\rangle = 6 k_BT
\gamma \delta _{nn'}\delta (t-t').
\label{eq10}
\end{equation}

The force $\vec{F}_{tot}(\vec{r}_n)$ includes all the forces resulting from the
potentials Eqs.~\ref{eq1}-\ref{eq3}, as well as the repulsive structureless
walls, which interact  with the polymer beads by means of the WCA-potential.

A special comment is in order with respect to the computation of the (osmotic)
pressure tensor, which is given by the Virial theorem as ($\rho$ is the density
of monomers in the system)

\begin{equation}
p_{\alpha \beta} = \rho k_BT \delta _{\alpha \beta} + \frac {1}{3V} \langle \sum
\limits _n r_n^\alpha F_{tot}^\beta (\vec{r}_n)\rangle ,
\label{eq11}
\end{equation}
where the sum is taken over all monomers of all chains, and it is important to
include the three-body forces resulting from the bond bending potential in
computing the total force acting on the $n$'th bead~\cite{milchev15}. This 
matters particularly when one extends Eq.~\ref{eq11} to define the local 
pressure tensor $p_{\alpha \beta} (z)$ in the interval $[z,z+dz]$ at distance 
$z$ from the planar wall. This local pressure tensor is needed for the 
computation of the surface tension of the polymer solution due to the 
wall~\cite{irving50,rowlinson82,todd95},
\begin{equation}
\gamma_{wall} = \frac 1 2 \int \limits _0^{L_z} dz [p_{zz}(z) - \frac 1 2
(p_{xx}(z)
+p_{yy}(z))] 
\label{eq12}
\end{equation}

\begin{figure}[htb]
\vspace{0.7cm}
\includegraphics[scale=0.33]{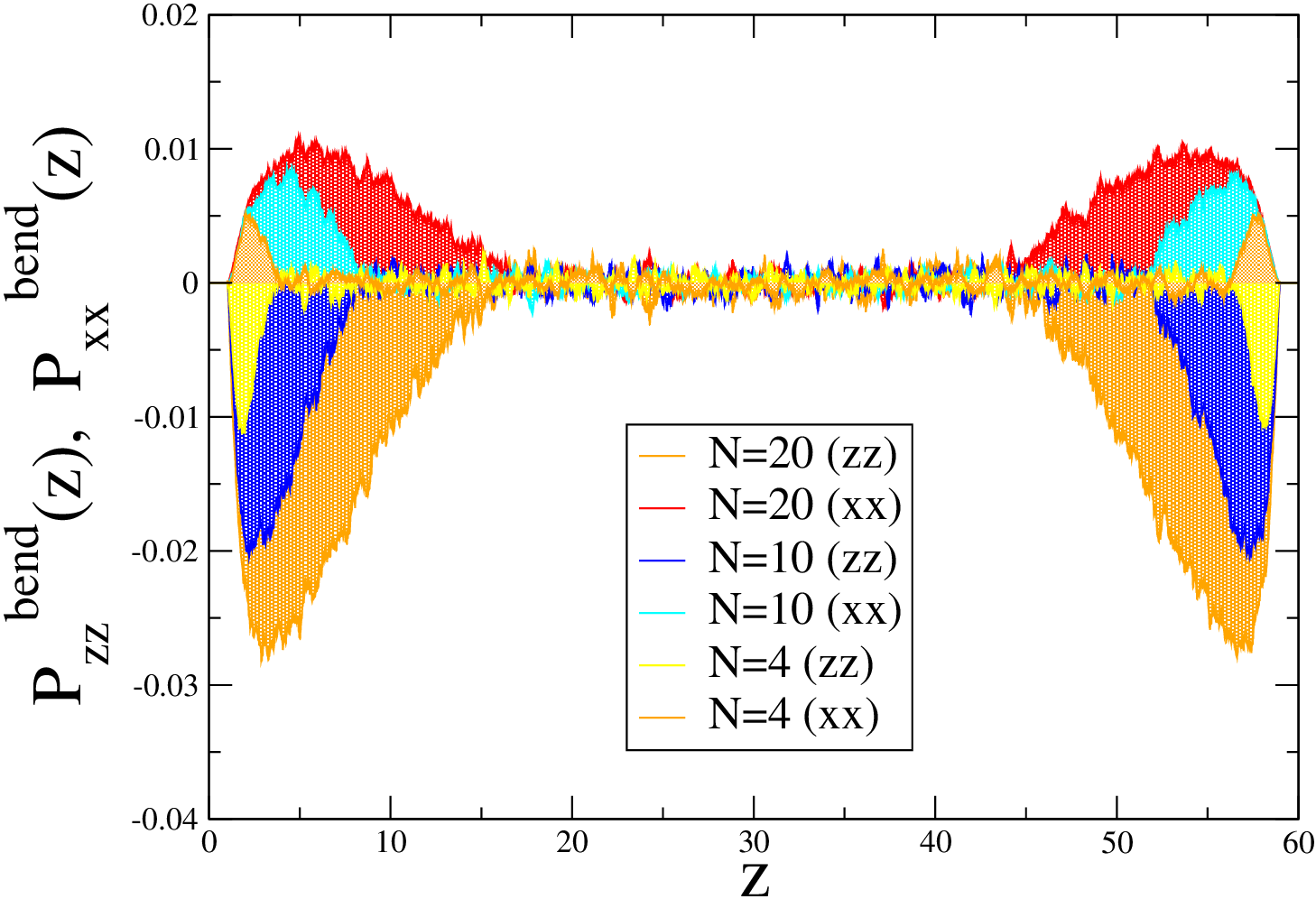}

\caption{
Contributions of the bending potential to
the normal and parallel components of the pressure tensor $p_{zz}^{bend}(z),\;
p_{xx}^{bend}(z)$ across a slit of width $L_z=60$ for chains with different 
length $N=4,\; 10$ and $20$,  monomer density $\rho = 0.0625$, and bending 
stiffness $\epsilon_b/k_BT=100$. } 
\label{fig1}
\end{figure}

In this equation, it has been anticipated that we do not deal with a semi-infinite system bounded by one repulsive wall in practice in a
simulation, but rather 
one deals with a thin film of height $L_z$, bounded by two equivalent walls 
having a surface area $L_{box} \times L_{box}$ each (in the $x,y$ directions 
parallel to these walls, periodic boundary conditions are used). Thus, the 
anisotropy of the pressure tensor includes the surface tension from both 
(equivalent) walls. As an example, Fig.~\ref{fig1} shows the contributions from 
the bending potential to $p_{zz}(z)$ and to $p_T(z)= (p_{xx}(z)+p_{yy}(z))/2$ 
for rather stiff polymers with $\epsilon_b/k_BT=100$. It is seen that that for 
the case of rather stiff polymers, the anisotropic contribution from the 
three-body forces is significant over a distance $z \approx R_g \approx aN$ 
adjacent to the walls. In the bulk where no direction is singled out these 
contributions always cancel, irrespective of stiffness. However, for very stiff 
chains, for which $R_g \ge L_z/2$, the anisotropy effects spread out essentially 
over the whole film, and the use of Eq.~\ref{eq12} would become unreliable; 
Eq.~\ref{eq12} implies that a well-defined separation of the pressure tensor 
into bulk and surface terms is possible, and this requires that the bulk 
behavior actually can be observed in the center of the film.

\subsection{Some Remarks About Our Implementation of Density Functional Theory
(DFT)} \label{subsection2c}

In our implementation of DFT, we use a slightly different microscopic model from
the one described in Section~\ref{subsection2a}. Specifically, instead of
employing FENE and WCA pairwise segment-segment potentials (Eqs.~(\ref{eq1}) and
(\ref{eq2})), we model the polymer molecule as a necklace of tangent hard
spheres of diameter  $\sigma$, whose degree of stiffness is governed by the
bending potential given by Eq.~(\ref{eq3}).

The starting point of any DFT approach~\cite{evans79} is the expression for the
Helmholtz free energy as a functional of the nonuniform molecular density
$\rho_{mol}(\vec{r},\omega)=\rho_{iso}(\vec{r})f(\vec{r},\omega)$, where
$\rho_{iso}(\vec{r})$ is the isotropic (orientationally independent) part and
$f(\vec{r},\omega)$ is the spatially- and angularly-dependent orientational
distribution function. 
Here $\omega$ is a compact notation for the polar angles 
$\theta$ and $\phi$, and $f(\vec{r},\omega)$ is defined as an average over all the
bonds~\cite{cao06c} and is normalized according to: 

\begin{equation}
\int d\omega f(\vec{r},\omega)=
\int_{0}^{2\pi} d\phi \int_{0}^{\pi} d\theta \sin\theta f(\vec{r},\theta,\phi)=1. 
\label{eqnorm}
\end{equation}

For the isotropic phase in the bulk,
$f_{iso}(\omega)=1/(4\pi)$ everywhere. However, even though we will be primarily
concerned with isotropic regime in the present work, due to the presence of
confining flat walls, which exert an ordering effect on the polymers,
$f(\vec{r},\omega)$ is expected to display a non-trivial spatial and angular
variation.

In this regard, we note that there have been several earlier DFT-based studies
of semiflexible polymers confined in a flat
slit.~\cite{forsman03,forsman06,turesson07} However, in these works only the
{\em orientationally independent} molecular density $\rho_{iso}(\vec{r})$ was
considered. At the same time, as will be seen below, even in the isotropic state
the nontrivial orientational dependence (due to the walls) plays an important
role in determining the surface tension $\gamma$. On the other hand, the
existing DFT-based studies of isotropic-nematic behavior of semiflexible
polymers in the {\em bulk}~\cite{fynewever98,cao04,jaffer01} operate with the
{\em spatially uniform} molecular density
$\rho_{mol}(\vec{r},\omega)=\rho_{mol}f(\omega)$, where $\rho_{mol}$ is the
molecular bulk number density.

For the present problem, where one needs to take into account {\em both} spatial
and angular dependence of $\rho_{mol}(\vec{r},\omega)$, it is necessary to
generalize the previously developed DFT approaches accordingly. While several
DFT-based studies of hard rods~\cite{vanroij00,cao06c} and
spherocylinders~\cite{delasheras03} confined in a slit have been reported, no
comparable method, to the best of our knowledge, has been proposed for
semiflexible molecules (note, that Chen and Cui~\cite{chen95b} used
self-consistent field theory (SCFT) to study the structure of the orientational
wetting layer of semiflexible polymers in the vicinity of a hard-wall surface;
however, it has been established that DFT, in general, is more accurate than
SCFT in resolving fine structural details of polymers at a
wall~\cite{egorov10b}). Hence, the aim of the present work is to develop a
method capable of treating {\em both} angular anisotropy and spatial
inhomogeneity of semiflexible polymers within the DFT framework.

Quite generally, one can write the Helmholtz free energy functional as a sum of
the ideal and excess terms:
\begin{equation}
\frac{F(\rho_{mol}(\vec{r},\omega))}{k_BT}=
\frac{F_{id}(\rho_{mol}(\vec{r},\omega))}{k_BT}+
\frac{F_{exc}(\rho_{mol}(\vec{r},\omega))}{k_BT}.
\label{eqfhelm}
\end{equation}

The ideal term is known exactly:
\begin{equation}
\frac{F_{id}(\rho_{mol}(\vec{r},\omega))}{k_BT}=
\int d\vec{r} \int d\omega \rho_{mol}(\vec{r},\omega) 
(\ln[4\pi \rho_{mol}(\vec{r},\omega)]-1).
\label{eqfideal}
\end{equation}

The excess term we split into ``isotropic''
($F_{exc}^{iso}(\rho_{iso}(\vec{r}))$) and ``orientational''
($F_{exc}^{orient}(\rho_{mol}(\vec{r},\omega))$) components. The former, which
depends only on the isotropic part of the molecular density, is calculated from
the Generalized Flory Dimer (GFD) theory,~\cite{honnell89} as described in
detail in Refs.~\onlinecite{forsman03,forsman06,turesson07}. The latter is
obtained on the basis of a density expansion (around the spatially and angularly
isotropic fluid) truncated at the second-order term:~\cite{patra97}
\begin{eqnarray}
\frac{F_{exc}^{orient}(\rho_{mol}(\vec{r},\omega))}{k_BT}
&=&
\int d\vec{r} \int d\omega \int d\vec{r}^{\prime} \int d\omega^{\prime}
a_{resc}^{PL}(\rho_{iso}(\vec{r}))
(\rho_{mol}(\vec{r},\omega)-\rho_{mol}/(4\pi))\nonumber \\ 
&\times&
(V_{exc}(\vec{r},\vec{r}^{\prime},\omega,\omega^{\prime})-\langle
V_{exc}^{iso}\rangle )
(\rho_{mol}(\vec{r}^{\prime},\omega^{\prime})-\rho_{mol}/(4\pi)),
\label{eqforient}
\end{eqnarray}

where $V_{exc}(\vec{r},\vec{r}^{\prime},\omega,\omega^{\prime})$ is the excluded
volume for 2 semiflexible polymers  with angular orientations $\omega$ and
$\omega^{\prime}$ (from which we have subtracted its spherical average $\langle
V_{exc}^{iso}\rangle$ in order to avoid the double-counting of the isotropic
contribution to the excess free energy, which is already taken into account via
the GFD-based term). In the above equation, $a_{resc}^{PL}(\rho_{iso}(\vec{r}))$
is the (spatially dependent~\cite{delasheras03})
Parsons-Lee~\cite{parsons79,lee87} rescaling factor, which is given in
Ref.~\onlinecite{delasheras03}. This   factor is needed in order to account for
the higher-order virial coefficients in this Onsager-like~\cite{onsager49}
expression for the excess free energy. The central quantity  in
Eq.~(\ref{eqforient}) is the spatially- and orientationally-dependent excluded
volume $V_{exc}(\vec{r},\vec{r}^{\prime},\omega,\omega^{\prime})$. While an
explicit analytical expression is known for this quantity for 2 rigid rods under
planar confinement,~\cite{shundyak01} no comparable expression is available for
2 semiflexible molecules. Accordingly, we adopt a simple decoupling
approximation and write
$V_{exc}(\vec{r},\vec{r}^{\prime},\omega,\omega^{\prime})\approx\delta(\vec{r}
-\vec{r}^{\prime})V_{exc}(\omega,\omega^{\prime})$, where for the
angularly-dependent term $V_{exc}(\omega,\omega^{\prime})$ we use an empirical
expression  due to Fynewever and Yethiraj obtained by fitting the corresponding
two-chain simulation data.~\cite{fynewever98} The corresponding spherical
average is given by $\langle V_{exc}^{iso}\rangle=\int d\omega \int
d\omega^{\prime} V_{exc}(\omega,\omega^{\prime})/(16\pi^2)$. 
In what follows, we will study inhomogeneous semiflexible polymer solution
confined by two infinite flat hard walls located at $z=0$ and $z=L_z$.
Accordingly, the isotropic molecular density profile is a function of $z$ only
and the corresponding expression for the grand potential takes the form: 
\begin{equation}
\Omega(\rho_{mol}(z,\omega)) = F(\rho_{mol}(z,\omega))
+\int_{0}^{h}dz \int d\omega \rho_{mol}(z,\omega)[V_{ext}^{mol}(z,\omega)-\mu],
\label{eqgrand}
\end{equation}
where $\mu$ is the polymer chemical potential and $V_{ext}^{mol}(z,\omega)$ is 
the external potential due to the two hard walls acting on the polymer 
molecules.

The equilibrium distributions $\rho_{iso}(z)$ and $f(z,\omega)$ are obtained by 
minimizing the grand potential with respect to $\rho_{iso}(z)$ and 
$f(z,\omega)$, respectively~\cite{lichtner12}. In practice, the minimization is 
performed in two steps~\cite{telodagama84}. First, one minimizes $\Omega$ with 
respect to $\rho_{iso}(z)$ as described in detail in 
Refs.~\onlinecite{forsman03,forsman06,turesson07}, which yields the following 
result for the equilibrium isotropic 
distribution~\cite{forsman03,forsman06,turesson07}: 

\begin{eqnarray}
\rho_{iso}(z)
&=& 
e^{\mu/k_BT}\sum_{i=1}^{N}\int_{0}^{L_z}\delta(z-z_i)\prod_{j=1}^{N}e^{-\lambda(z_j)}
\prod_{k=1}^{N-1}\Theta(|\Delta z_k|-\sigma) \nonumber \\
&\times&
\prod_{l=1}^{N-2}\Psi(\Delta z_{l},\Delta z_{l+1})dz_1\cdots dz_N,
\label{rhoisoz}
\end{eqnarray}
where the indices $i,j,k,l$ label individual monomers, i.e. the total density 
distribution is written as a sum over monomer density distributions.

In the above equation,
\begin{eqnarray}
\Psi(\Delta z_{i},\Delta z_{i+1})
&=&
\exp[\epsilon_b/k_BT(1-\Delta z_{i}\Delta z_{i+1}/\sigma^2)] \nonumber \\
&\times&
I_{0}\left\{\epsilon_b/k_BT\left[1-\left(\frac{\Delta 
z_{i}}{\sigma}\right)^2\right]^{1/2} \left[1-\left(\frac{\Delta 
z_{i+1}}{\sigma}\right)^2\right]^{1/2}\right\}, 
\label{psi} 
\end{eqnarray}
where $I_{0}(x)=\frac{1}{2\pi}\int_{0}^{2\pi}\exp[-x\cos\phi]d\phi$ is a 
modified Bessel function and $\Delta z_{i}= z_{i+1}-z_{i}$. In addition, 
$\lambda(z)=[\delta F_{exc}^{iso}/\delta \rho_{iso}(z)+v_{ext}^{mon}(z)]/k_BT$, 
where $v_{ext}^{mon}(z)$ is the external potential due to the two hard walls 
acting on individual monomers, which is equal to zero for $0<z<L_z$ and is 
infinite otherwise. Finally, $\Theta(x)$ in Eq.~(\ref{rhoisoz}) is the Heaviside 
step function, which is equal to unity for non-positive values of its argument 
and is equal to zero for $x>0$. 

Given that $\rho_{iso}(z)$ in Eq.~(\ref{rhoisoz}) is written as a sum of 
contributions from individual monomers, one can readily obtain the {\em zeroth} 
approximation to the orientational distribution function, $f_0(z,\omega)$, which 
includes the orienting effect on the polymer molecules due to the hard 
walls~\cite{cao06c}, but not due to the Onsager-like term in 
Eq.~(\ref{eqforient}), which has not been treated yet. 

As the second step in the minimization procedure, we now minimize the grand 
potential with respect to $f(z,\omega)$, which yields the following result for 
the equilibrium orientational distribution 
function:~\cite{lichtner12,telodagama84} 
\begin{equation}
f(z,\omega)=C(z)\exp[-V_{eff}(z,\omega)/k_BT]\exp[-2a_{resc}^{PL}(\rho_{iso}(z)) 
\int d\omega^{\prime}(\rho_{mol}(z,\omega^{\prime})-\rho_{mol}/(4\pi)) 
(V_{exc}(\omega,\omega^{\prime})-\langle V_{exc}^{iso}\rangle )],
\label{eqfomega}
\end{equation}
where $C(z)$ is the normalization constant ensuring that $\int d\omega
f(z,\omega)=1$ for all $z$. The angular- and spatially-dependent effective 
external potential is given by $V_{eff}(z,\omega)=-k_BT\ln[f_0(z,\omega)]$. In 
practice, Eq.~(\ref{eqfomega}) is solved iteratively, by using $f_0(z,\omega)$ 
as the initial guess and iterating until the converged result for $f(z,\omega)$ 
is obtained. Note that due to the averaging over the $xy$ (hard wall) plane, the 
orientational distribution function $f(z,\theta)$ depends on $z$ and $\theta$ 
only, but not on the azimuthal angle $\phi$, which precludes the treatment of 
biaxiality within our spatially one-dimensional DFT approach. 

Once the equilibrium orientational distribution function 
$\rho_{mol}(z,\theta)=\rho_{iso}(z)f(z,\theta)$ is computed, 
one can readily obtain the order parameter as a function of the distance from 
the wall: 
\begin{equation}
S(z)=2\pi \int_{0}^{\pi} d\theta f(z,\theta) 
(\frac{3}{2}\cos^2\theta-\frac{1}{2}). \label{orderparameter}
\end{equation}
Recall that the orientational distribution function $f(z,\omega)$ is defined as
an average over all the bonds in the molecule.~\cite{cao06c}  From the above
definition it is clear that $S(z)=0$ corresponds to random chain orientation,
while $S(z)=-0.5$ corresponds to perfect alignment of the chain parallel to the
wall. 

In addition, various thermodynamic quantities can be calculated, including the
surface tension at the wall. As follows from the above discussion, $\gamma$
contains isotropic and orientational contributions in both its ideal and excess
terms. For example, for the isotropic contribution to the ideal term one
gets:~\cite{forsman03,forsman06,turesson07}
\begin{equation}
\frac{\gamma_{id}^{iso}\sigma^2}{k_BT}=\int_{0}^{h}dz(\rho_{mol}-\rho_{iso}(z)),
\label{gammaidiso}
\end{equation}
and for the orientational contribution one obtains:~\cite{perera88}
\begin{equation}
\frac{\gamma_{id}^{orient}\sigma^2}{k_BT}=2\pi\int_{0}^{h}dz\int_{0}^{\pi} 
d\theta \rho_{mol}(z,\theta)\ln[4\pi\rho_{mol}(z,\theta)/\rho_{mol}].
\label{gammaidorient}
\end{equation}
Likewise, the expressions for the isotropic and orientational contributions to
the excess part of the surface tension can be readily obtained from the
corresponding parts of the excess Helmholtz free energy.

In presenting the DFT results below, we will split the total surface tension
into its isotropic and orientational components:
$\gamma=\gamma^{iso}+\gamma^{orient}$, where 
$\gamma^{iso}=\gamma^{iso}_{id}+\gamma^{iso}_{exc}$ and 
$\gamma^{orient}=\gamma^{orient}_{id}+\gamma^{orient}_{exc}$. 

\section{Molecular Dynamics Results for Semiflexible Polymers at Repulsive Walls}
\label{section3}

In presenting our results in this and the following section, we make all 
distances dimensionless by measuring them in units of the size parameter 
$\sigma$ and all energies in units of the thermal energy $k_BT$. 

\begin{figure}[htb]
\vspace{1.0cm}
\includegraphics[scale=0.3, angle=0]{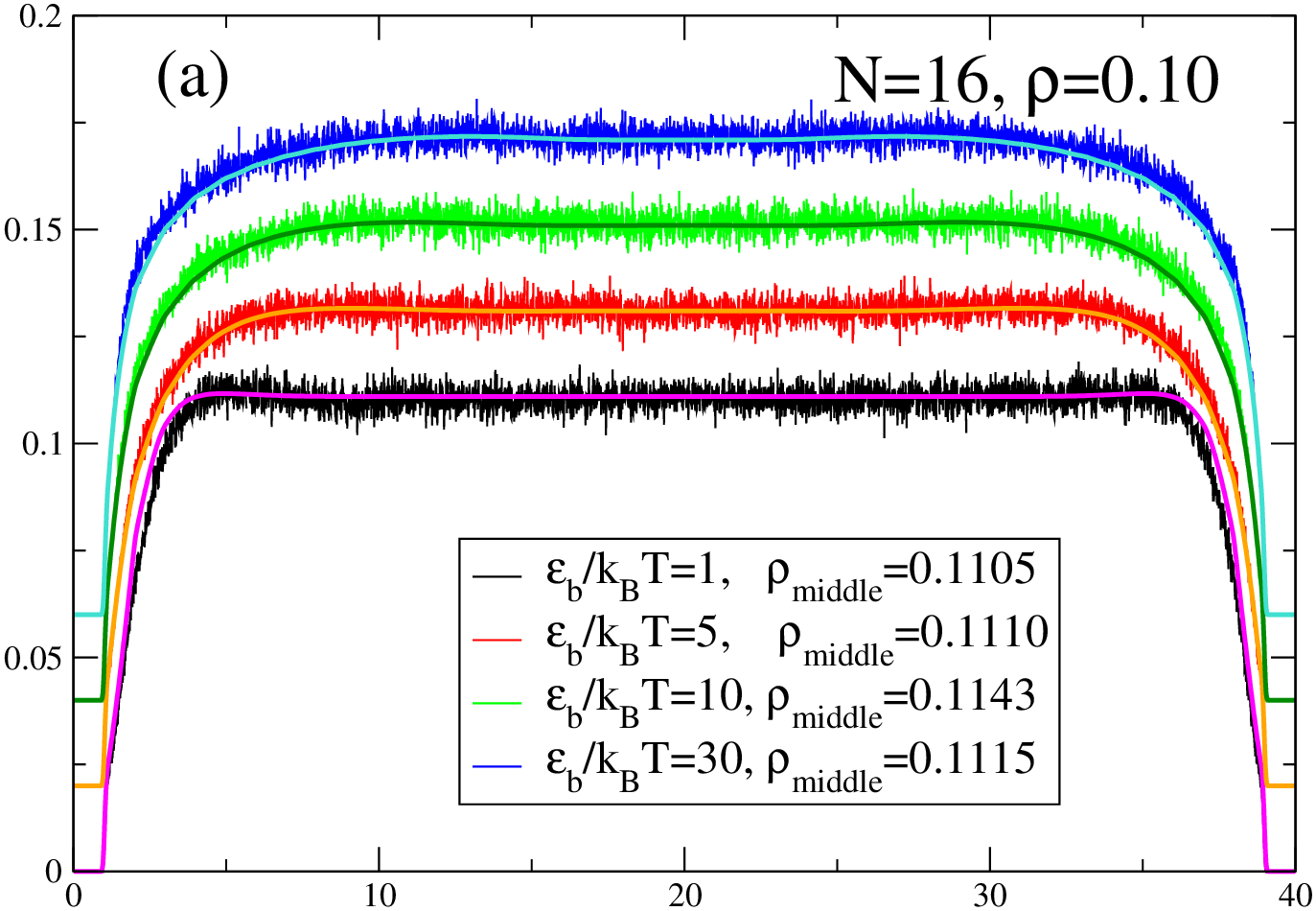}
\hspace{0.9cm}
\includegraphics[scale=0.3, angle=0]{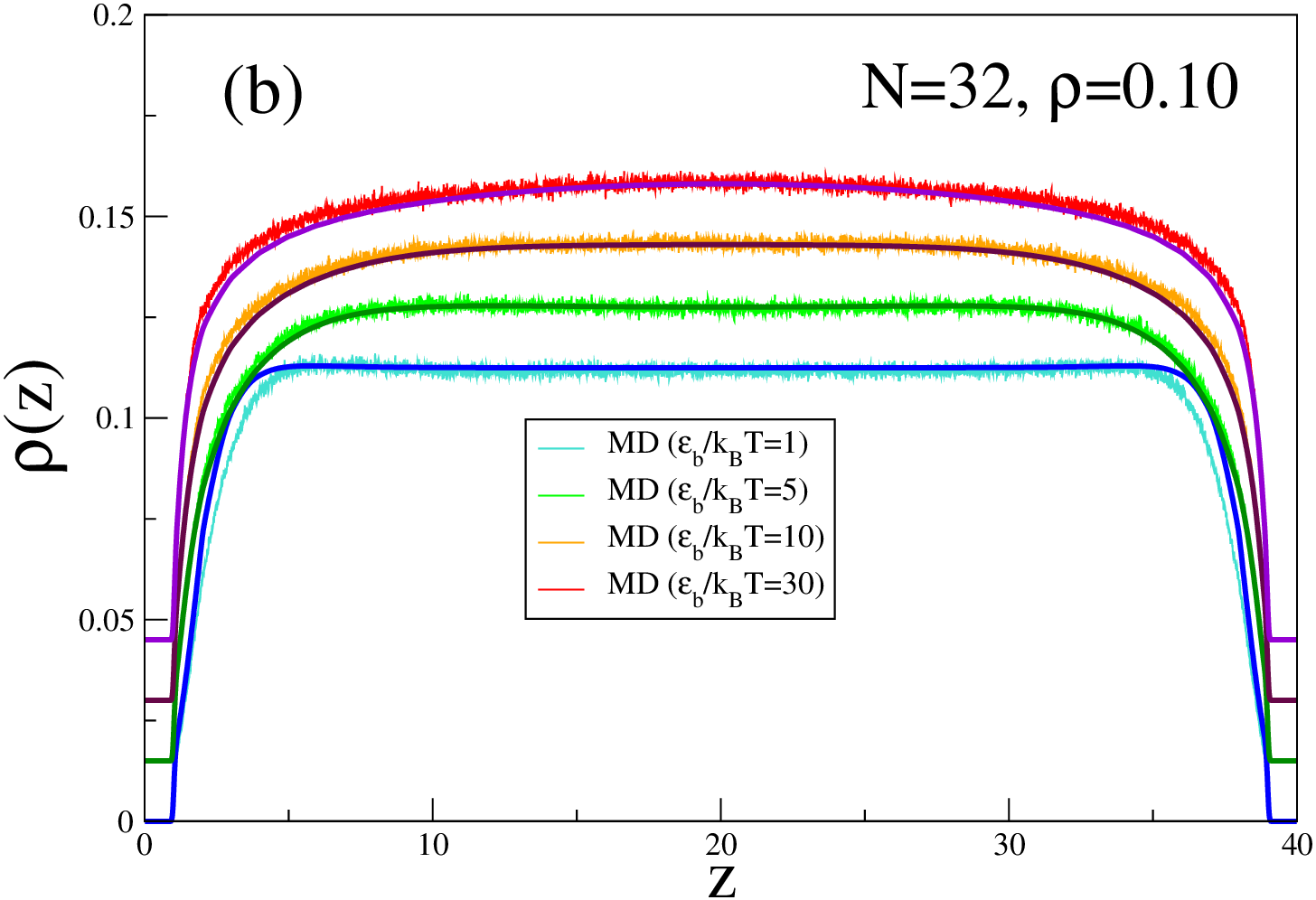}
\vspace{0.3cm}

\caption{(a) The monomer density profiles $\rho(z)$ across the film for $L_z=40$ 
and four values of the chain stiffness parameter, $\epsilon_b/k_BT=1,5,10$ and 
$30$, for $N=16$ and $\rho = 0.1$. Noisy lines are from MD, smooth lines from 
DFT (performed at the values of the bulk density $\rho_b$ corresponding to the 
MD values of $\rho_{middle}$ indicated in the legend). All profiles are 
shifted vertically by $0.02$ for better visibility. (b) The monomer density 
profiles $\rho(z)$ across the film for $L_z=40$ and four values of the chain 
stiffness parameter, $\epsilon_b/k_BT=1,\;5,\;10$ and $30$, for $N=32$ and
$\rho_b = \rho_{middle}$. Noisy lines are from MD, smooth lines from DFT, and
the profiles are shifted vertically by $0.015$.}
\label{fig2}
\end{figure}

In Fig.~\ref{fig2} we display the impact of growing chain stiffness on the
distribution of monomer density across a slit with $L_z = 40$ for a system of
semiflexible polymers at concentration $\rho=0.1$ and two different chain
lengths, $N=16,\;N=32$. One may detect a qualitative change in the density
profiles whereby an increasingly pronounced depletion of macromolecules in the
vicinity of the confining walls is observed irrespective of chain length $N$.
As a consequence, the density sufficiently far away from the walls exceeds
slightly yet steadily the average density in the slit with increasing rigidity,
$\epsilon_b/k_BT=1,\; 5,\; 10,\; 30$, which should be kept in mind when MD data
are compared to DFT results where density corresponds to that in a grand
canonical ensemble. 
Nonetheless, Fig.~\ref{fig2} manifests a very good agreement between the two
methods, MD and DFT, as far as the profiles of monomer density are concerned,
whereby the DFT results have been obtained for $\rho_b = \rho_{middle}$. 
When the density profile gets horizontal over an extended range of $z$ near the middle point $z_{middle}=L_z/2$ of the slit, the two walls are essentially independent of each other, and $\rho_{middle}$ should be equal to the bulk density of a semi-infinite system. 



\begin{figure}[htb]
\vspace{0.9cm}
\begin{center}
\includegraphics[scale=0.3]{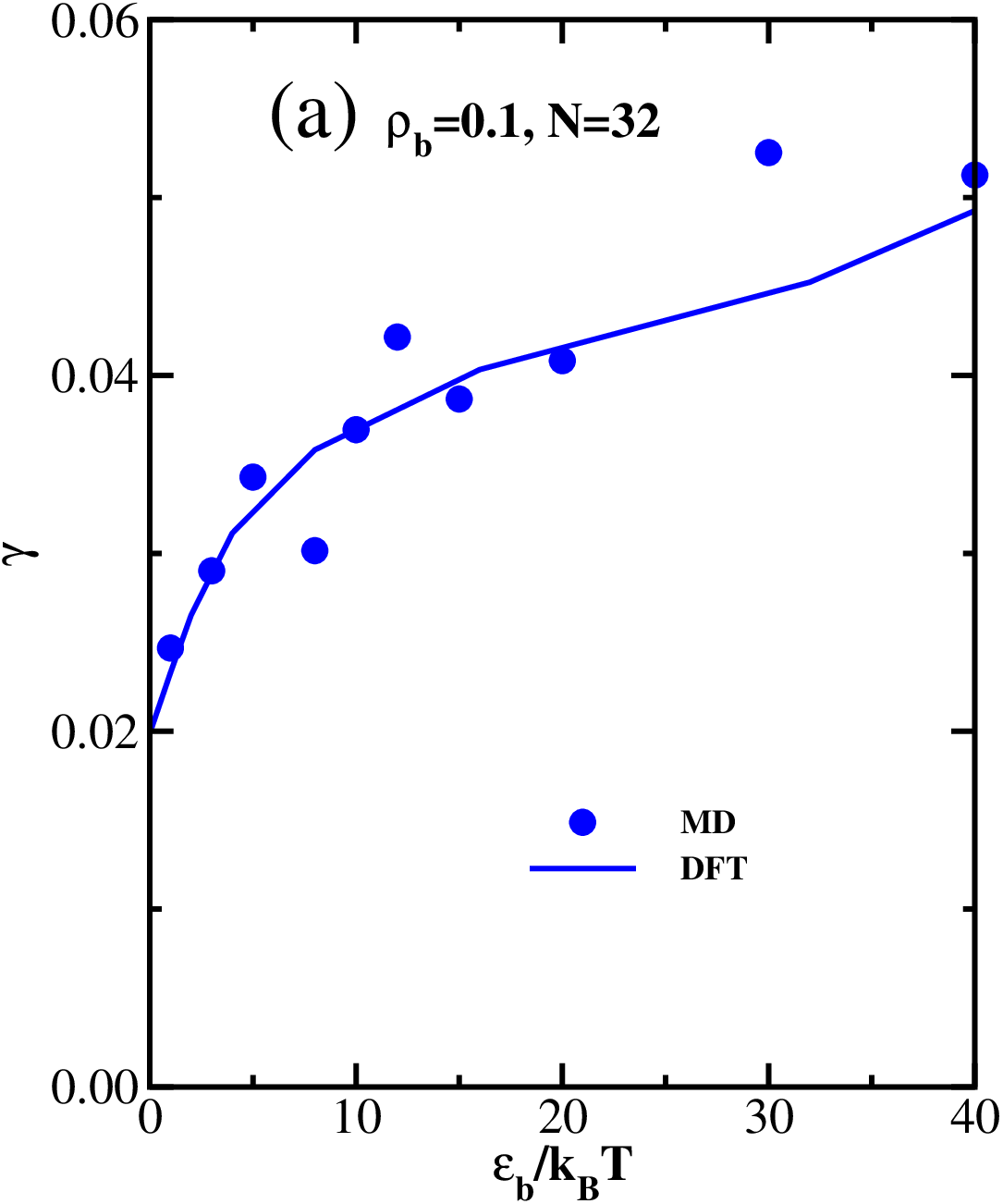}
\hspace{0.5cm}
\includegraphics[scale=0.26]{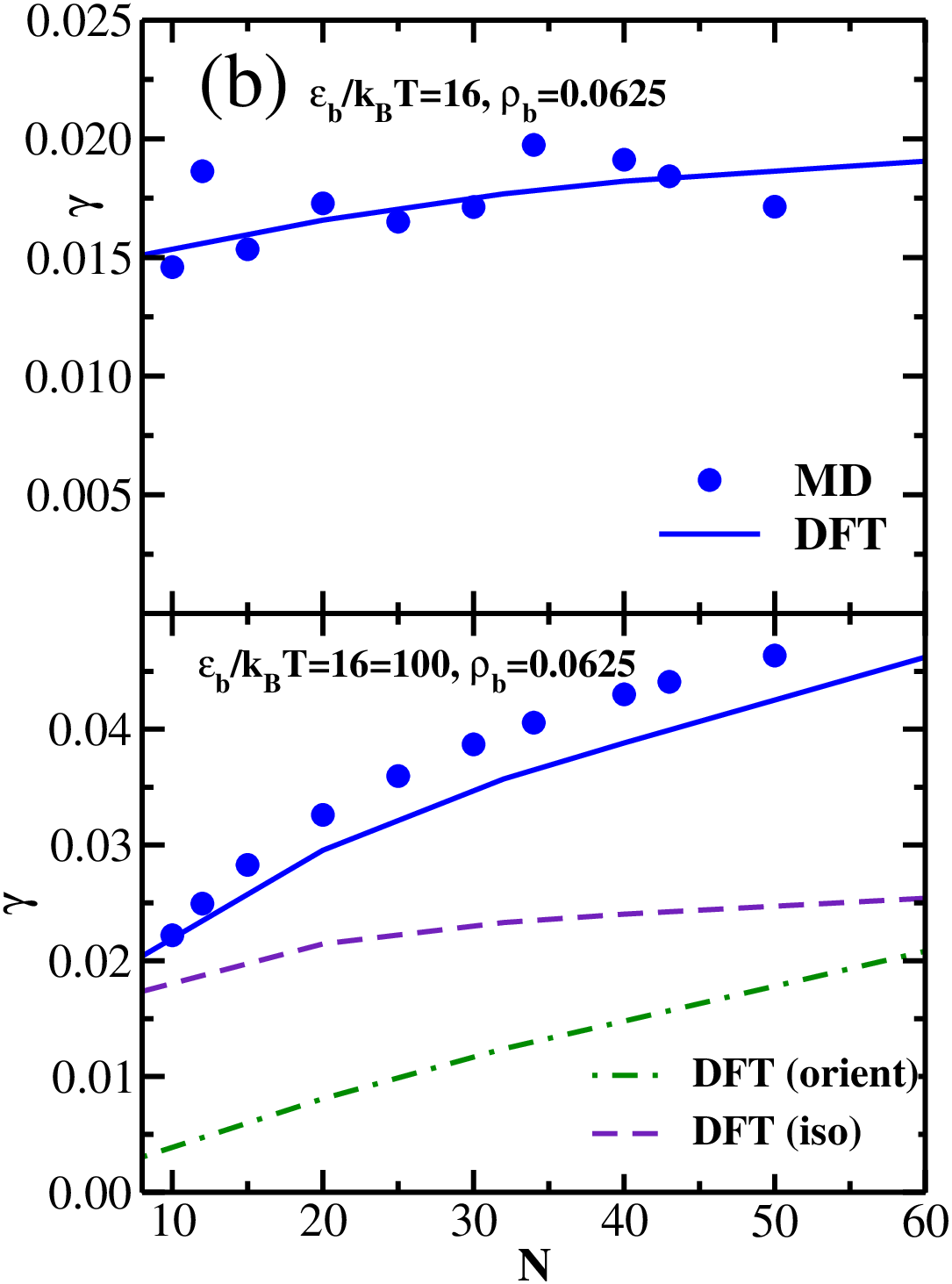}

\end{center}
\caption{(a) Surface tension as a function of the chain stiffness parameter 
$\epsilon_b/k_BT$ for $\rho_b=0.1$, $N=32$, comparing MD results (dots) with DFT 
predictions (lines). (b) Surface tension as a function of the chain length for 
$\rho_b=0.065$  and the stiffness parameter $\epsilon_b/k_BT = 16$ (upper panel) 
and $\epsilon_b/k_BT = 100$ (lower panel), comparing MD results (dots) with DFT 
predictions (lines); the decomposition of the DFT result into isotropic and 
orientational components is shown for the stiff chain, 
$\epsilon_b/k_BT = 100$.}
\label{fig3}
\end{figure}

Next we focus on the behavior of the surface tensions $\gamma_{wall}$ in the
regime of densities $\rho \leq 0.1$ and not extremely stiff chains, so we stay
far away from the isotropic to nematic transition (Fig.~\ref{fig3}). At these
low densities, accurate estimation of the osmotic pressure tensor components
spatially resolved near repulsive walls is rather difficult, and hence the
application of Eq.~\ref{eq12} suffers from rather large statistical errors. In
\begin{figure}[htb]
\vspace{0.5cm}
\includegraphics[scale=0.3, angle=0]{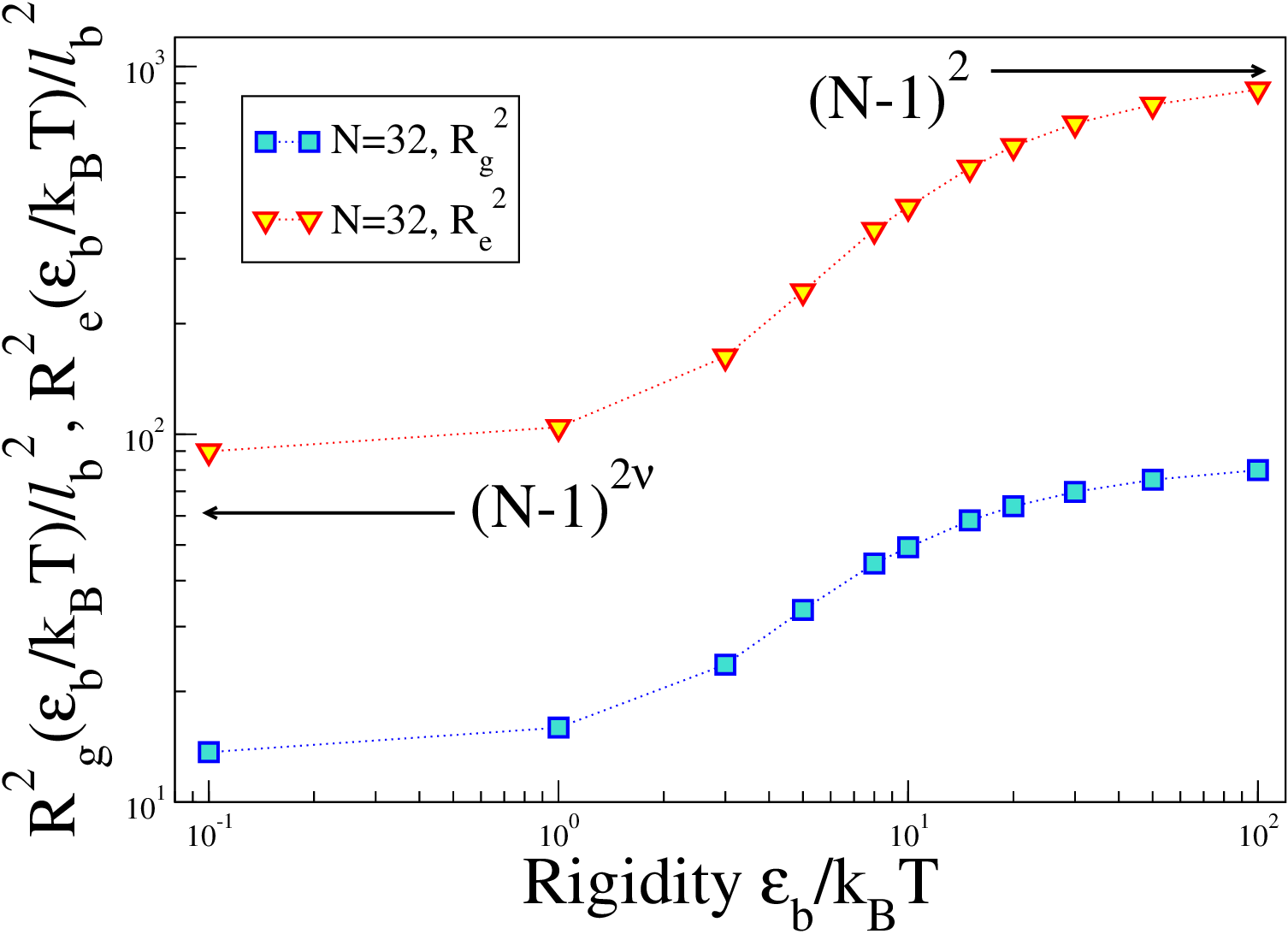}

\caption{Mean square end-to-end distance $\langle R_e^2\rangle$ and mean square 
gyration $\langle R_g^2\rangle$ radius (plotted in units of the squared bond 
length $\ell_b$) vs. the rigidity parameter $\epsilon_b/k_BT$. All data were 
taken for a density $\rho_b=0.0625$ in a box of linear dimensions $40 \times 40 
\times 40$, using periodic boundary conditions and a chain length $N=32$. 
Arrows denote the squared end-to-end distance $R_e^2/\ell_b^2$ in the limit of 
a flexible coil and a rigid rod.}
\label{fig4}
\end{figure}
Fig.~\ref{fig3}a we present the surface tension as a function of the chain 
stiffness parameter $\epsilon_b/k_BT$ for $\rho_b=0.1$, $N=32$, comparing MD 
results (dots) with DFT predictions (lines). One sees that the two methods are 
in nearly quantitative agreement, both predicting monotonic increase of the 
surface tension with increasing $\epsilon_b/k_BT$. In Fig.~\ref{fig3}b we 
present the surface tension as a function of the chain length for $\rho_b=0.065$ 
 for two values of the stiffness parameter: $\epsilon_b/k_BT=16$ (upper panel) 
and $\epsilon_b/k_BT = 100$ (lower panel). Once again, MD and DFT are in nearly 
quantitative agreement. For more flexible chain ($\epsilon_b/k_BT = 16$), both 
methods predict that $\gamma$ increases very slowly with $N$, while for the 
stiffer chain ($\epsilon_b/k_BT =100$), the increase is more pronounced. For the 
stiffer chain, we show the decomposition of the DFT result for $\gamma$ into the 
isotropic and orientational terms; one sees that the orientational term becomes 
increasingly prominent with increasing chain length. For more flexible chain, 
the DFT result for $\gamma$ is dominated by the isotropic term (decomposition 
not shown). The smallness of the surface tension in this region of parameters is 
expected, of course, due to the smallness of the considered density: with 
increasing density the surface tension increases rather fast. Note that due to 
the large fluctuations of the MD data in Fig.~\ref{fig3} we have disregarded the 
distinction between the average density $\rho$ in the simulation box and the 
bulk density $\rho_b$ (which is seen only near $z \approx h/2$ and only if $L_z$ 
is chosen large enough, cf. Fig.~\ref{fig1}).


\begin{figure}[htb]
\vspace{0.7cm}
\includegraphics[scale=0.3, angle=0]{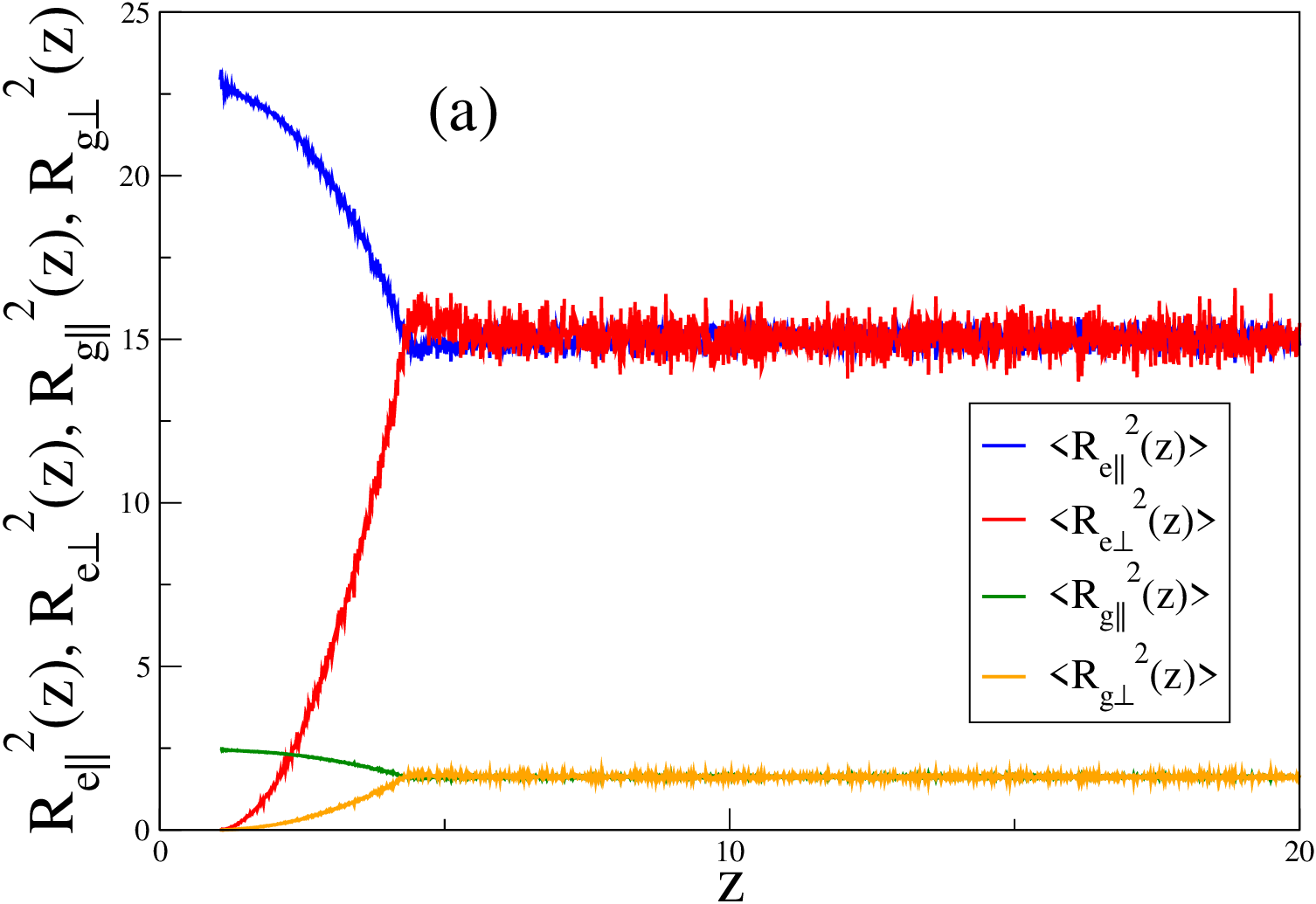}
\hspace{0.3cm}
\includegraphics[scale=0.3, angle=0]{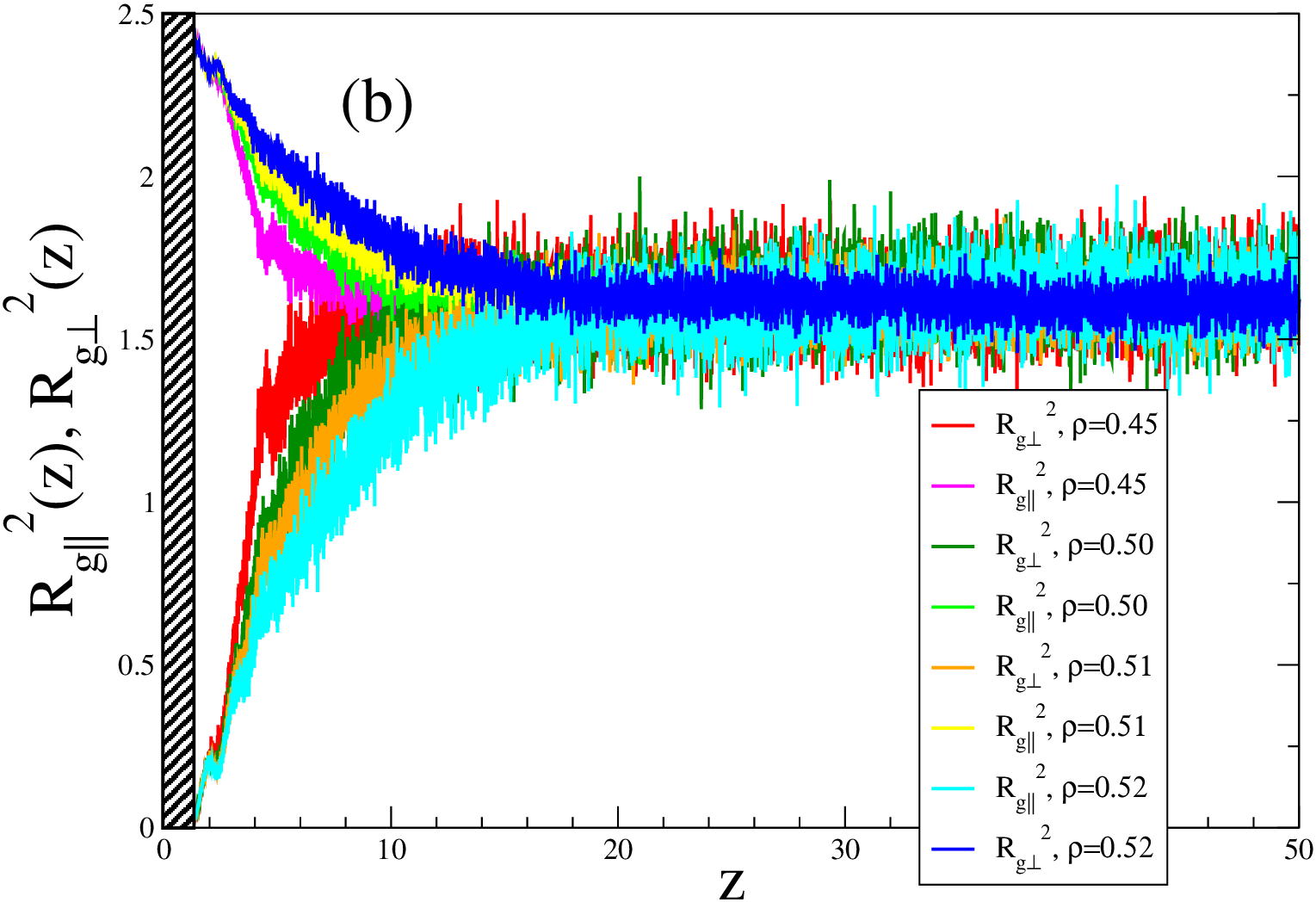}
\hspace{0.3cm}
\vspace{1.cm}

\includegraphics[scale=0.32, angle=0]{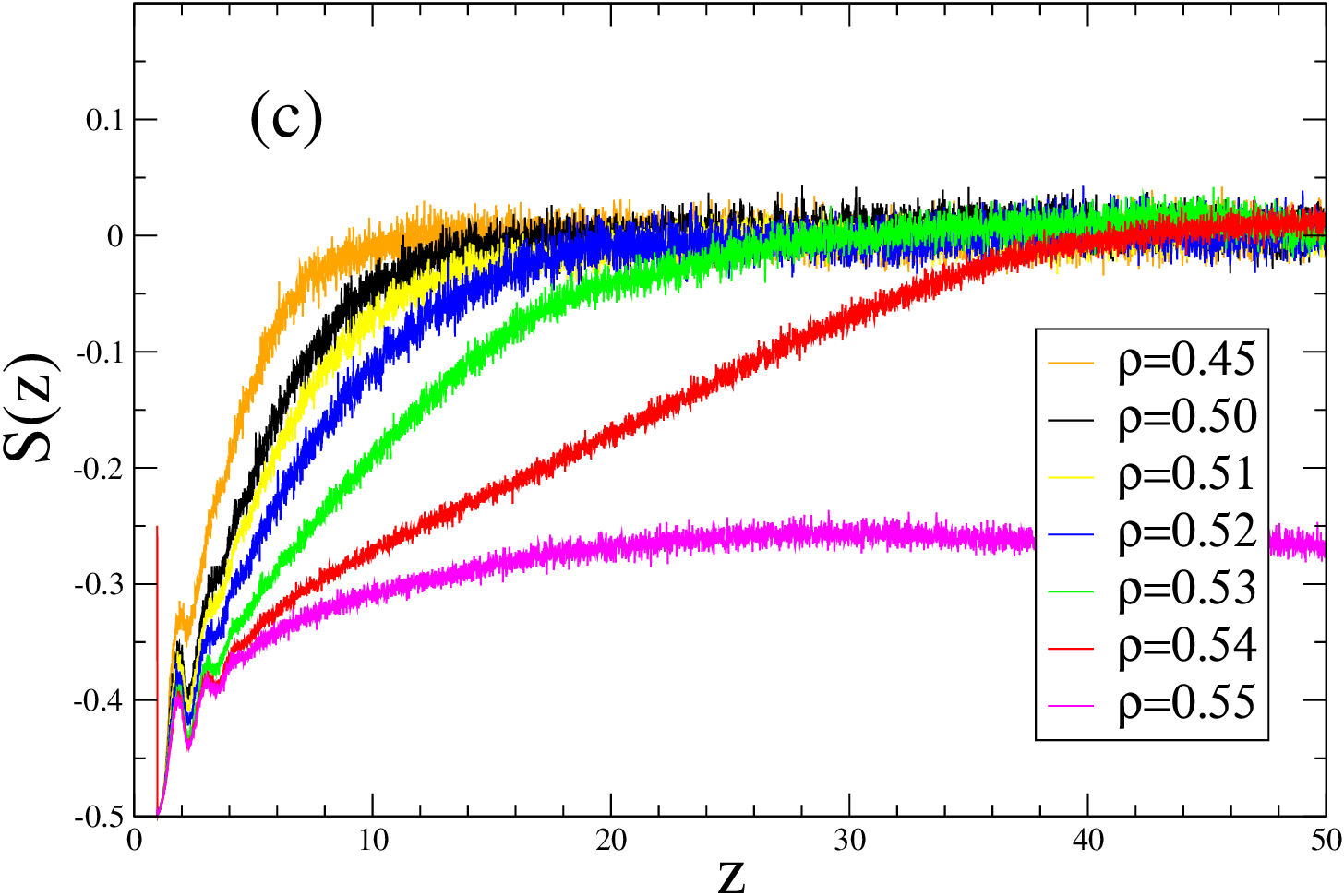}
\hspace{0.3cm}
\includegraphics[scale=0.3, angle=0]{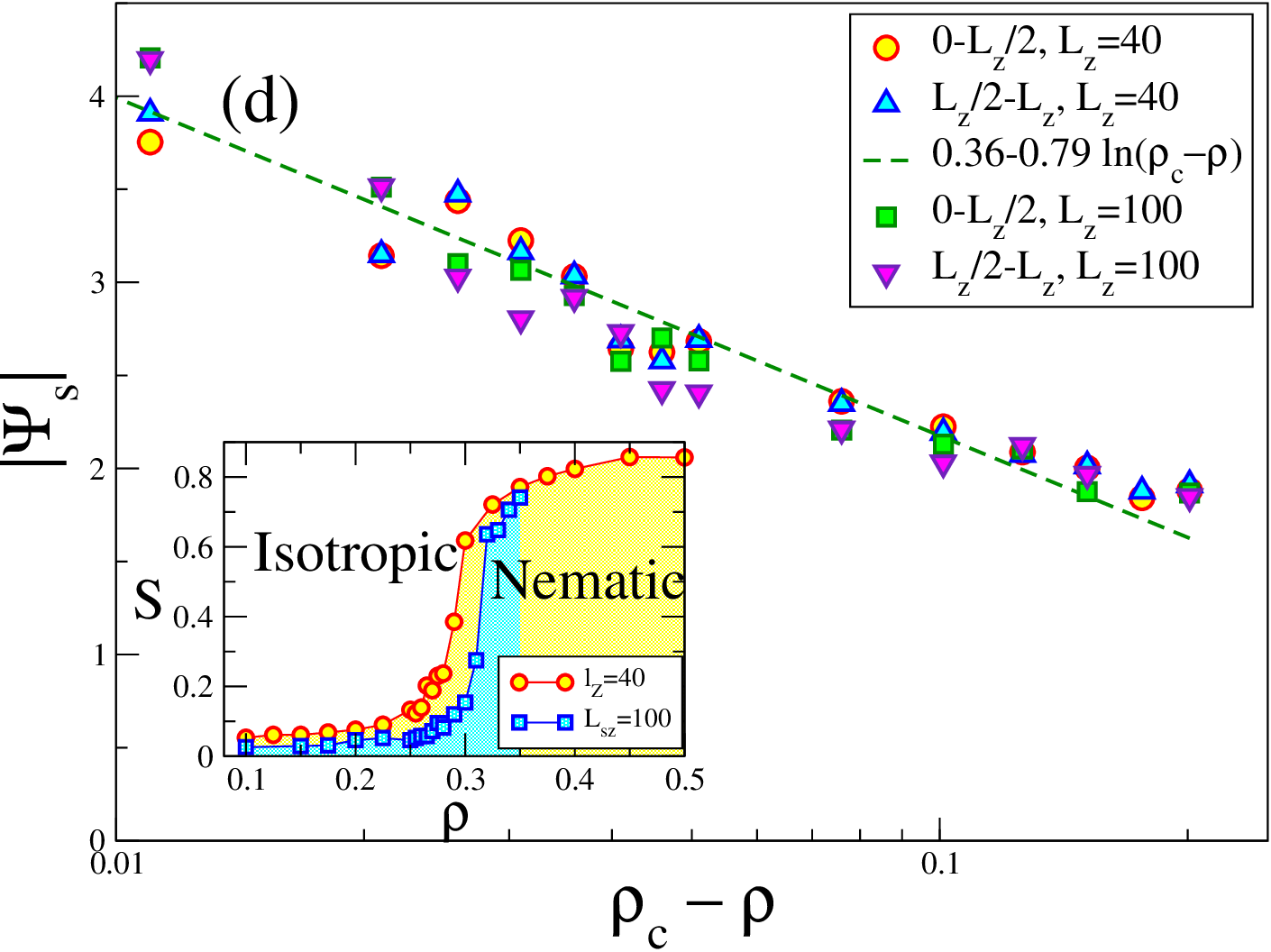}

\caption{(a) Components of the mean square end to end distance $\langle 
R_e^2(z)\rangle$ and radius of gyration $\langle R^2_g(z)\rangle$, parallel 
$(\langle R_{e||}^2(z)\rangle),\langle R_{g||}^2(z)\rangle$ and perpendicular 
$(\langle R_{e\bot}^2(z)\rangle,\langle R_{g\bot}^2(z)\rangle)$ to the confining 
walls, plotted versus the distance $z$ of the center of mass of each polymer 
from the nearest wall. All data refer to $N=8$, $\epsilon_b/k_BT=100$, average 
density in the system being $\rho=0.1$, and $L_z=40$. (b) Components of the mean 
square gyration radius of the polymers parallel $(\langle R_{g||}^2(z)\rangle)$ 
and perpendicular $(\langle R_{g\bot}^2(z)\rangle)$ to the confining walls 
plotted as a function of distance $z$ of the center of mass of the polymer from 
the nearest wall. All data refer to $N=8$, $\epsilon_b/k_BT =100, L_z =100$, and 
several average densities in the simulation box are shown, as indicated in the 
key to the figure. (c) Mean orientation of bonds $S(z)$ [see text] plotted 
versus the distance $z$ of a bond from the nearest wall, for the case $N=8, 
\epsilon_b/k_BT=100, L_z = 100$, and various average densities in the simulation 
box, as indicated in the key to the figure. (d) Surface-induced excess order 
parameter $|\Psi_s|$, where $\Psi_s$ is obtained as $\Psi_s = 
\int_{0}^{L_z/2}dz S(z)$, plotted vs the density difference $\rho_c-\rho$, 
$\rho_c$ being the critical density where nematic order starts to set in. Note 
that the results for $2$ values of $L_z$ are included. Estimates are extracted 
from the surface-induced excess order parameter at both walls and shown 
individually, to indicate the magnitude of statistical errors. A logarithmic 
abscissa scale is used to show that the data are compatible with the expected 
logarithmic divergence as $\rho$ approaches $\rho_c$. The inset displays the 
phase diagram (i.e., the variation of $S$ with $\rho$) for the widths 
$L_z=40,\;100$. }
\label{fig5}
\end{figure}

While the variation of $\epsilon_b/k_BT$ on the surface tension has a rather weak
effect (Fig.~\ref{fig3}), one should recall that the change in the actual
conformations of the macromolecules is very pronounced (Fig.~\ref{fig4}). Under
the shown conditions one observes a crossover from self-avoiding walk-like
behavior $(\langle R_e^2\rangle \propto \ell_b^2N^{2 \nu})$ to rod-like behavior
$(\langle R_e^2\rangle \approx \ell_b^2(N-1)^2$ as $\epsilon_b/k_BT$ varies from
$\epsilon_b/k_BT = 0$ to $\epsilon_b/k_BT = 100$.
\begin{figure}[htb]
\vspace{0.7cm}
\includegraphics[scale=0.31, angle=0]{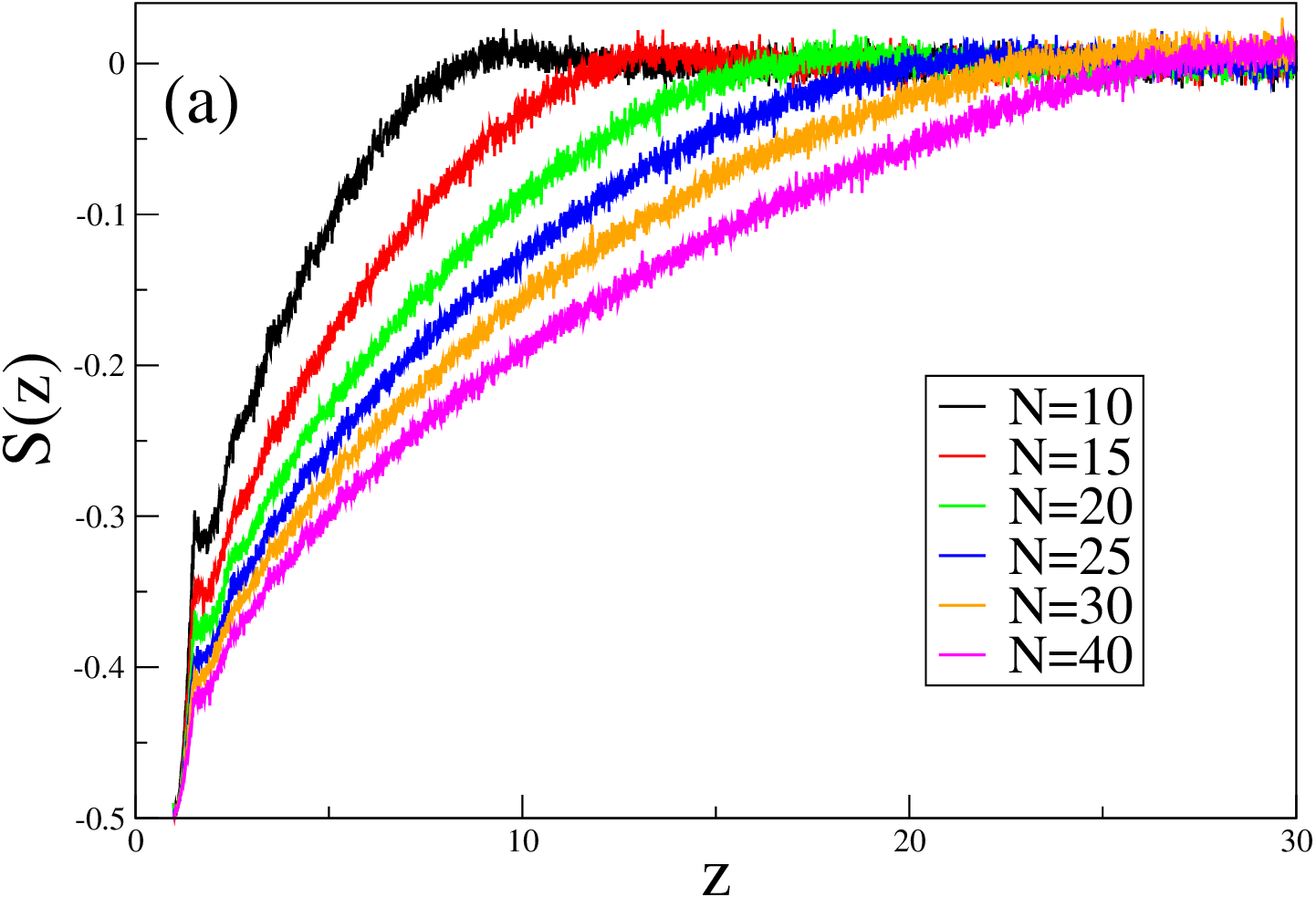}
\hspace{0.3cm}
\includegraphics[scale=0.31, angle=0]{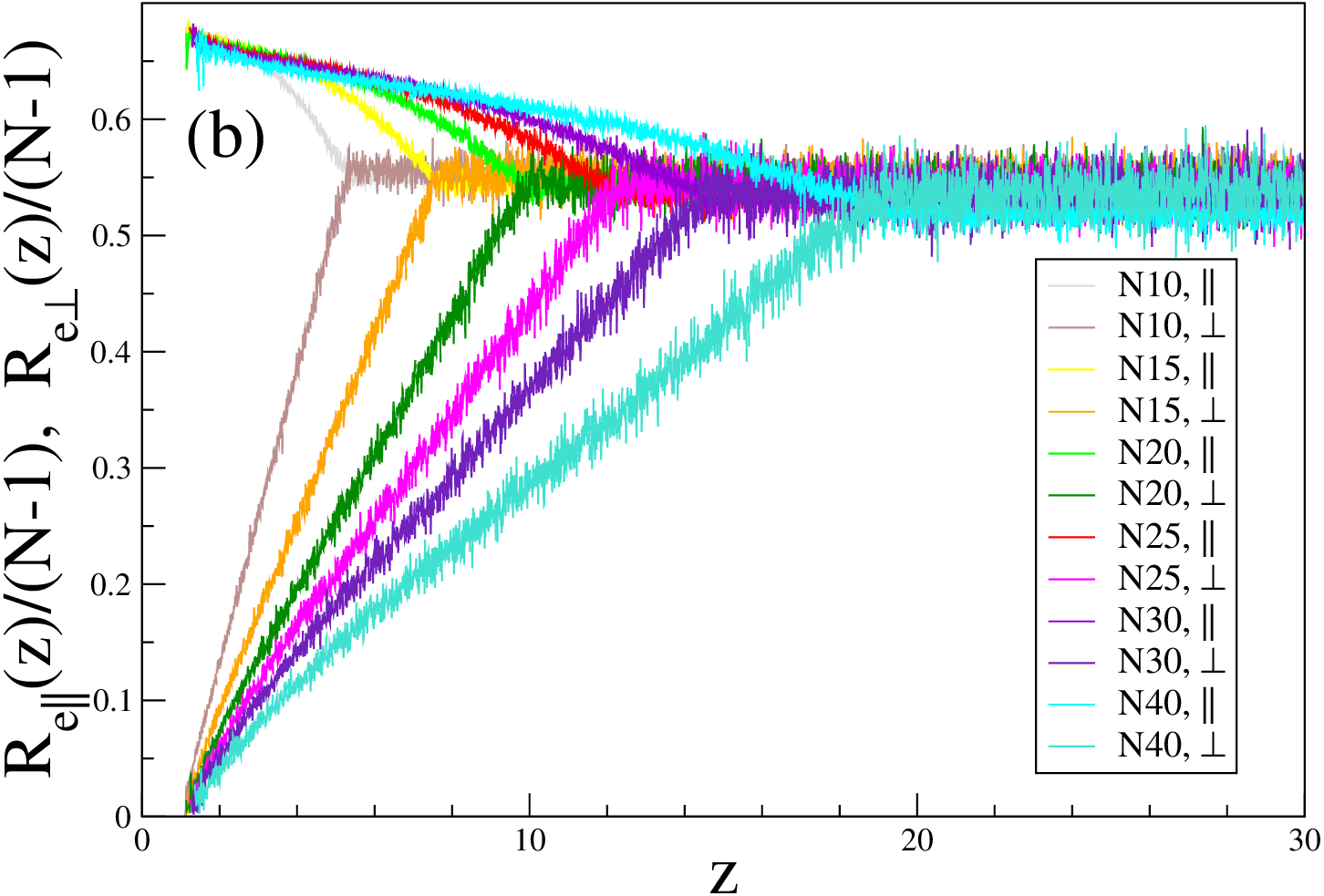}

\caption{(a) Bond orientational order parameter $S(z)$ plotted vs. the distance 
$z$ from the nearest wall, as in Fig.~\ref{fig5}c, but now for the case 
$\rho$=0.0625, $\epsilon_b/k_BT$=100, $L_z$=60, and chain lengths $N$ from 
$N=10$ to $N=40$, as indicated in the key to the figure. Note that $L_x = L_y = 
100$ and bins of thickness $\Delta z =0.01$ were chosen, to reach both good 
statistics and a fine enough resolution of the profile. (b) Components of the 
normalized end-to-end vector $(\langle R_{e\|}^2(z)\rangle)^{1/2}/(N-1)$,
$\langle R_{e\bot}^2(z)\rangle^{1/2}/(N-1)$, parallel and perpendicular to the
confining walls, plotted versus the distance $z$ of the center of mass of a
polymer from the nearest wall. Chain lengths $N$ from $N=10$ to $N=40$ are shown,
as indicated by the key. All data are for $\rho=0.0625, L_x = L_y=100$,
$L_z=60$, and $\epsilon_b/k_BT = 100$ whereby due to symmetry only the left half
of the profiles is shown.}
\label{fig6}
\end{figure}

Of course, it is very interesting to study the linear dimensions of the 
macromolecules when they are confined by the parallel repulsive walls. 
Fig.~\ref{fig5}a presents plots of the components of the mean square end-to-end 
distance parallel $(R_{e\|}^2(z))$ and perpendicular $(R_{e\bot}^2(z))$ to the 
wall, as well as the corresponding mean square gyration radii $(R_{g||}^2(z), 
R_{g\bot}^2(z))$, resolved as a function of distance $z$ of the center of mass 
of the chain, for a very short $(N=8)$ and stiff $(\epsilon_b/k_BT=100)$ 
polymer, at a small density $\rho = 0.1$. One sees that near the walls $(z\leq 
4)$ the perpendicular component is reduced and the parallel component is 
enhanced, indicating that the short stiff chains are rather strongly aligned 
parallel to the walls. The bulk value $R^2_{ez} = R_{\ell||}^2 \approx 15$ is 
only slightly smaller than the theoretical value for a rigid rod containing 
$N=8$ beads connected by links of length $\ell_b \approx 0.96, L^2/3 = 15.05$, 
and the value $R_{\ell ||}^2 \approx 22.5$ is consistent with the expected value 
$L^2/2$ for rigid rods of length $L=6.72$. For these stiff short chains the bulk 
ratio $R_e^2/R_g^2 \approx 10$ still falls below its limit (12) reached for long 
rods $N\rightarrow \infty$, as expected.

Closer to the transition isotropic/nematic in the bulk (Fig.~\ref{fig5}b), the
orienting effect of the wall on the short stiff chains extends much further than
half their length, and reflects the formation of the wall-induced nematic layer,
which can also be seen from the mean orientation of bonds (Fig.~\ref{fig5}c), as
measured by the second Legendre polynomial $S(z)=P_{2,z}(\cos \theta)=[3\langle
\cos ^2 \theta\rangle - 1]/2$, $\theta$ being the angle with respect to the
$z$-axis, normal to the confining walls. Recall that $S(z)=-0.5$ means perfect
alignment parallel to the walls. In Fig.~\ref{fig5}c, the $z$-coordinate of a
bond between monomers $i$ and $i+1$ ($i$ being an index labeling the monomers
along the considered chain) is simply defined as $z=(z_{i}+z_{i+1})/2$, and the
angular brackets $\langle\cdot\rangle$ denote an average over all the bonds of
all the chains that fall in an interval $[z-0.01,z+0.01]$. Since in isotropic
phase in the bulk the order parameter $S$ is equal to zero, one can define the
surface-induced excess order parameter as follows: $\Psi_s=\int_{0}^{L_z/2}dz
S(z)$; the corresponding results are presented in Fig.~\ref{fig5}d.

When the density $\rho$  of the effective monomers increases up to the critical
density $\rho_c$ where nematic order starts to occur in the bulk, a
surface-induced nematically ordered layer forms at the repulsive wall.
Fig.~\ref{fig5}d shows that the thickness of this nematic surface layer diverges
logarithmically towards infinity when $\rho$ tends towards $\rho_c$. 
A related surface-induced ordering has already been found for a lattice 
model~\cite{ivanov13,ivanov14}. 

Of course, it is also interesting to ask what changes when the length of the
polymers is varied. Fig.~\ref{fig6}a shows that even at a low average density
$\rho$ in the simulation box (recall that $\rho$ slightly differs from $\rho_b$,
as pointed out in the discussion of Fig.~\ref{fig1}), the range over which the
wall leads to predominantly parallel bond orientation increases substantially,
as $N$ increases. Note that for $\rho = 0.0625$ and $N\geq 30$ there is no
longer a well-defined extended bulk region of the isotropic phase (where
$P_2(\cos \theta) = 0$) for $L_z=60$. This is not evident from the components of
the end-to-end vector, however (Fig.~\ref{fig6}b). There the range over which
the wall strongly matters always seems to be simply $z \approx (N-1)\ell_b/2$.
But although for $N=40$ the components parallel and perpendicular to the wall
reach horizontal plateaus in the center, the fact that these plateaus differ
also is a clear evidence that there is no longer any bulk region in the system.
Note that despite the low density the behavior observed in Fig.~6 is very different from wall effects on a dilute solution of flexible polymers (which would behave like self-avoiding walks under good solvent conditions), but rather chains here are like slightly flexible rods, for the chosen parameters.

\section{DFT Results for Semiflexible Polymers at Repulsive Walls}
\label{section4}
\subsection{The effect of varying chain stiffness} 
\label{subsection4a}

We begin by considering the effect of varying the chain stiffness parameter on
the density profiles, bond orientational order parameter, and the surface
 \begin{figure}[htb]
\vspace{0.7cm}
\includegraphics[scale=0.31, angle=0]{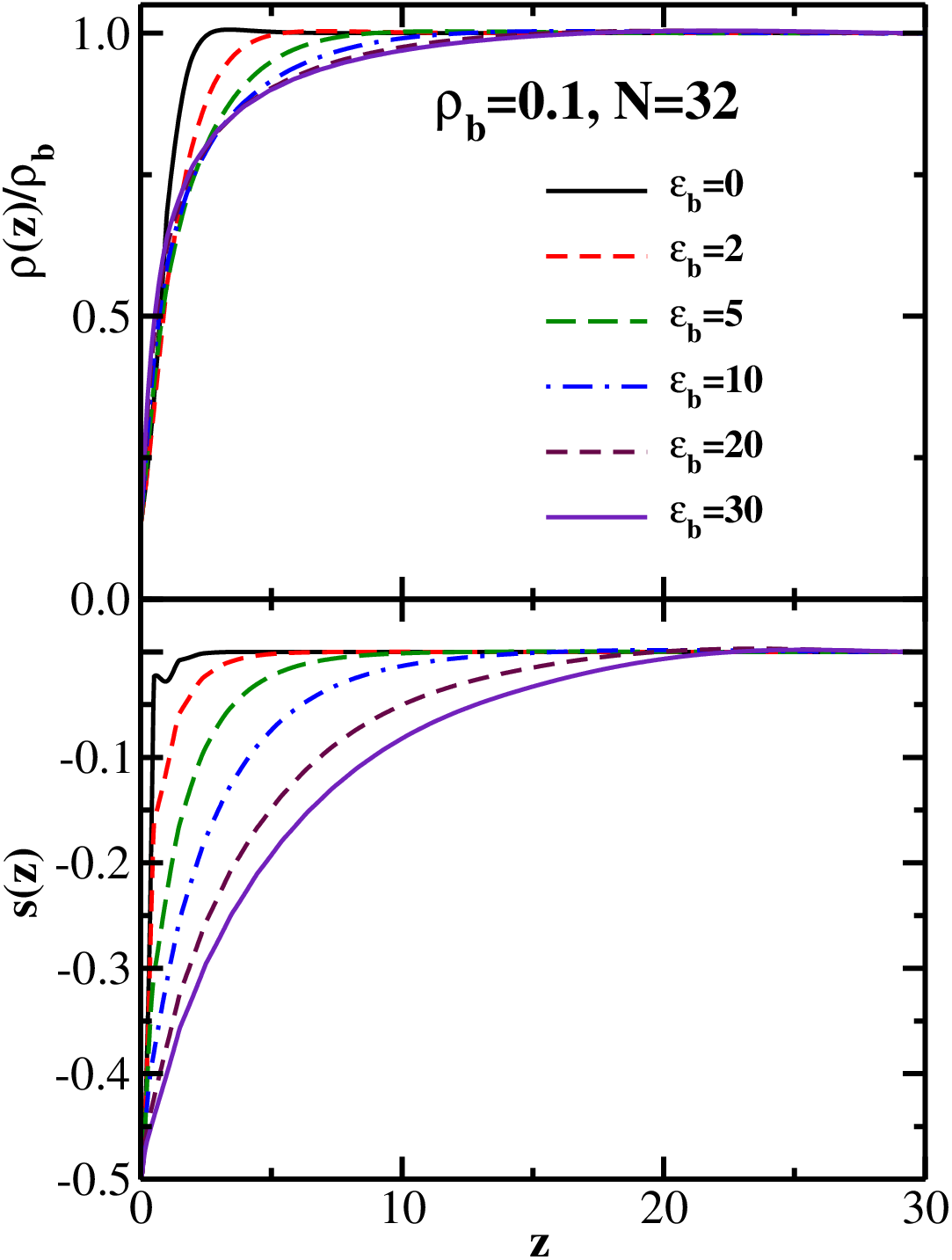}
\hspace{0.9cm}
\includegraphics[scale=0.31, angle=0]{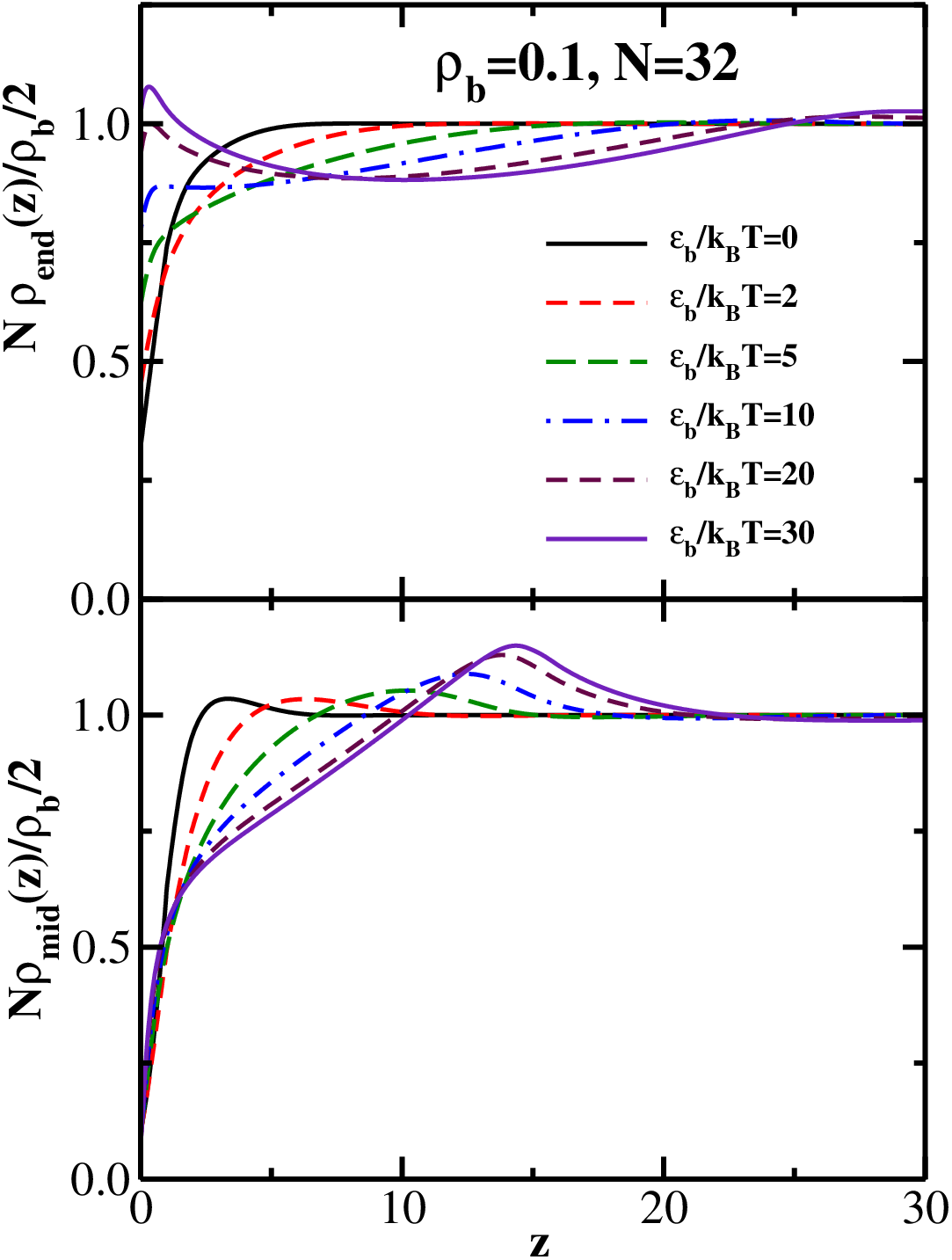}
\vspace{0.3cm}
\caption{(a) Upper panel: the normalized monomer density profiles 
$\rho(z)/\rho_{b}$ for six values of the chain stiffness parameter: 
$\epsilon_{b}/k_BT=$0, 2, 5, 10, 20, and 30; $N$=32 and $\rho_{b}$=0.1. Lower 
panel: same for the bond orientational order parameter $S(z)$. (b) Upper panel: 
the normalized end-monomer density profiles for six values of the chain 
stiffness parameter: $\epsilon_{b}/k_BT$=0, 2, 5, 10, 20, and 30; $N$=32 and 
$\rho_{b}$=0.1. Lower panel: same for the two middle monomers (segments number 
16 and 17).}
\label{fig7}
\end{figure}
tension. We set the chain length to $N$=32 and the bulk monomer density to 
$\rho_b$=0.1. The DFT results for the total monomer density profiles (normalized 
by the bulk monomer density) are shown in the upper panel of Fig.~\ref{fig7}a 
for several values of the chain stiffness parameter $\epsilon_{b}/k_BT$. As one 
would expect, the range of the depletion zone grows with increasing 
$\epsilon_{b}/k_BT$, which leads to increasing surface tension as was already 
seen in Fig.~\ref{fig3}. In the lower panel of Fig.~\ref{fig7}a we show DFT 
results for the bond orientational order parameter $S(z)$ defined by 
Eq.~(\ref{orderparameter}). One sees that with increasing chain stiffness  the 
tendency of the chains to be aligned parallel to the wall extends to larger 
distances from the wall. Returning to the monomer density profiles displayed in 
the upper panel of Fig.~\ref{fig7}b, one observes an (almost imperceptible) 
maximum in these profiles beyond the depletion zone. The same phenomenon has 
been reported in an earlier study of fully flexible chains, where the appearance 
of this maximum was related to the segregation of end-monomers to the 
wall.~\cite{shvets13}

\begin{figure}[htb]
\vspace{0.7cm}
\includegraphics[scale=0.31, angle=0]{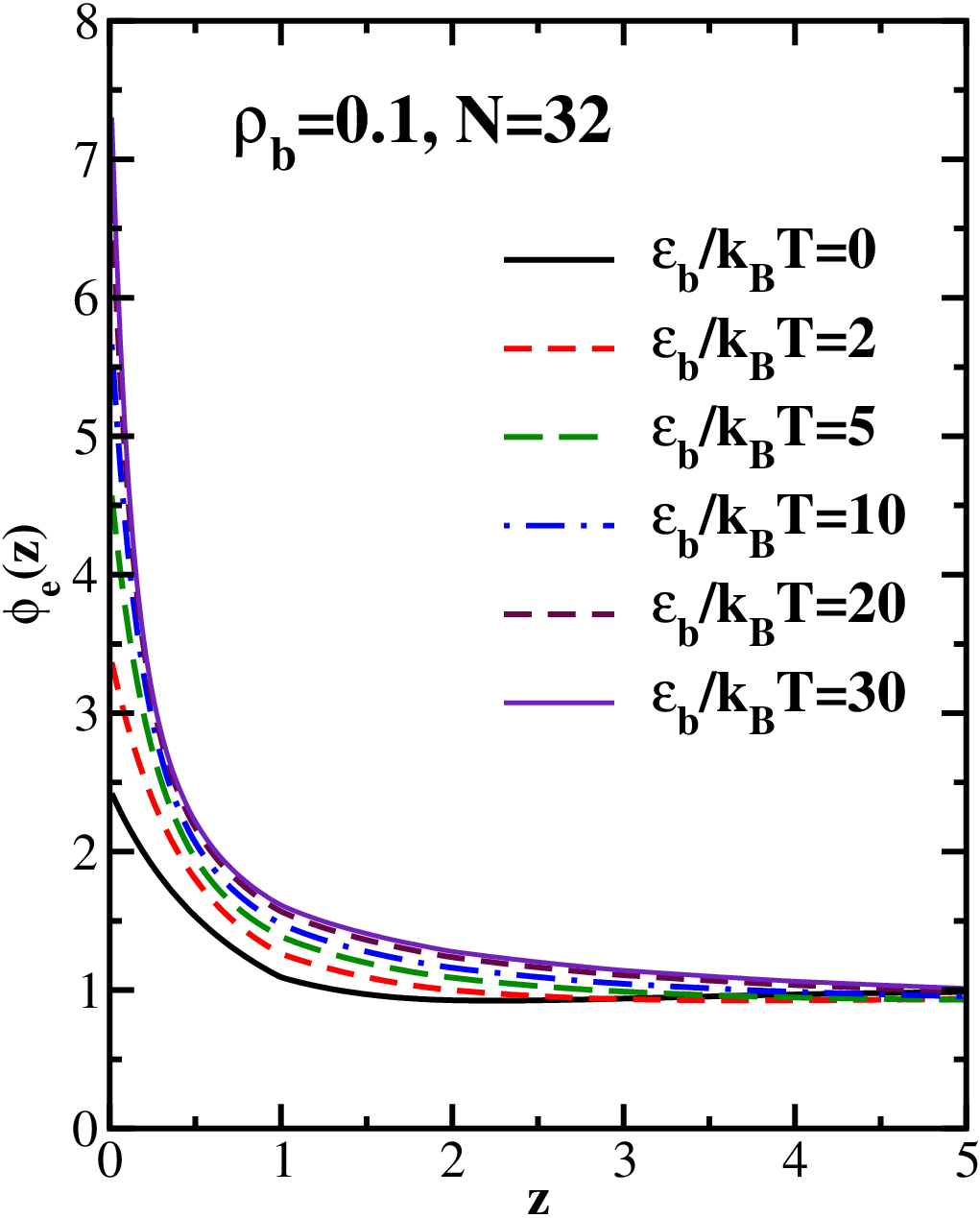}
\hspace{0.9cm}
\includegraphics[scale=0.31, angle=0]{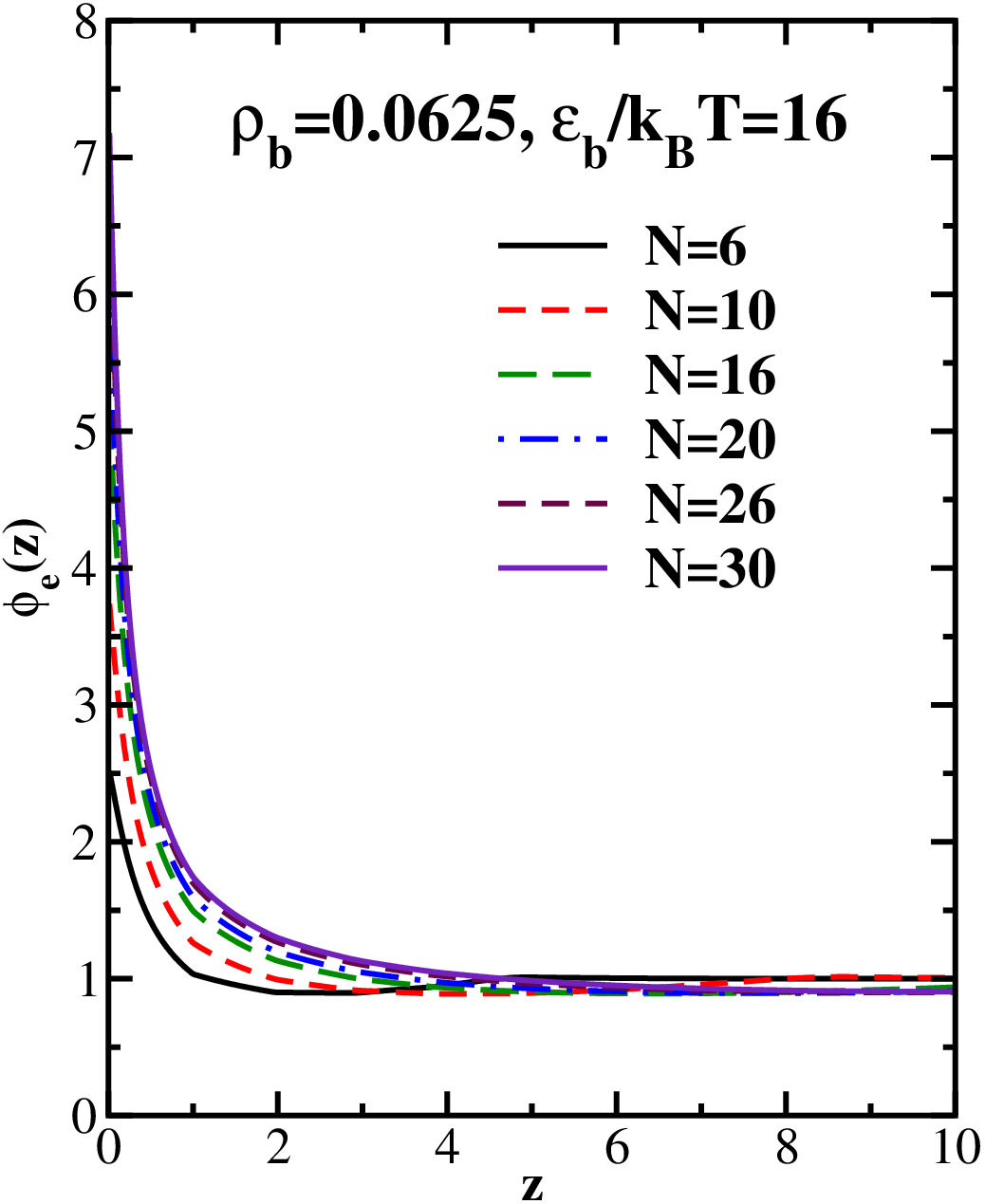}
\vspace{0.3cm}
\caption{(a) The ratio of the end-monomer to the total density profile for six 
values of the chain stiffness parameter: $\epsilon_{b}/k_BT=$0, 2, 5, 10, 20, 
and 30; $N=32$, and $\rho_{b}$=0.1. (b) The ratio of the end-monomer to the 
total density profile for six values of the chain length: $N$=6, 10, 16, 20, 26, 
and 30; $\epsilon_{b}/k_BT$=16 and $\rho_{b}$=0.0625.} 
\label{fig8}
\end{figure}

Accordingly, it is of interest to consider the effect of the chain stiffness on 
the end-monomer density distributions. Indeed, the enrichment of chain ends at 
surfaces and interfaces has been studied for a long time in polymer 
melts~\cite{wang91c,matsen14}, while dilute polymer solutions have received 
little attention in this regard. In the upper panel of Fig.~\ref{fig7}b, we 
display the DFT results for the end-monomer density profiles $\rho_{end}(z)$ 
(normalized by their bulk values). One immediately sees that the contact value 
$\rho_{end}(0)$ grows dramatically with increasing $\epsilon_{b}/k_BT$. By 
contrast, the contact value of the total monomer density $\rho(0)$ (see upper 
panel of Fig.~\ref{fig7}a) is essentially $\epsilon_{b}/k_BT$-independent.
Accordingly, the ratio $\rho_{end}(0)/\rho(0)$ is expected to be a strongly
increasing function of $\epsilon_{b}/k_BT$. This is indeed confirmed in
Fig.~\ref{fig8}a where we plot the ratio of the end-monomer to the total
density profile defined by:~\cite{wang91c} 
\begin{equation}
\phi_e(z)=\frac{N}{2}\frac{\rho_{end}(z)}{\rho(z)}
\label{phiend}
\end{equation}
One sees that in the vicinity of the wall the end-monomer density is always 
enhanced relative to the total density (and the degree of this enrichment grows 
with $\epsilon_{b}/k_BT$), while away from the wall there is concomitant 
depletion (as the difference between the two normalized profiles must integrate 
to 0 over the entire slit). It is also worth pointing out that the absolute 
values of this enrichment at the wall are significantly greater compared to the 
case of polymer melt.~\cite{wang91c} Note also that the enrichment of chain ends at the walls is pronounced in a very narrow region (of width $\Delta z < \sigma$), while the corresponding adjacent depletion zone is spread out over a much broader region 
(of width $\Delta z \gg \sigma$), and hence is difficult to recognize visually in 
Figs.~8 and 9.  At this point we recall that MD simulations are performed in the canonical ensemble at the average density
$\rho$, while  DFT calculations are performed in the grand-canonical ensemble at
the bulk density $\rho_b$. 
Due to the depletion of the density near the walls in a slit of finite width, such as used in MD simulations, it is a nontrivial task (subject to both statistical and systematic errors) to convert the average density $\rho$ in MD to the corresponding bulk density $\rho_b$, although for the cases shown here we expect that $\rho$ and $\rho_b$ differ only slightly. 
 
\begin{figure}[htb]
\vspace{0.7cm}
\includegraphics[scale=0.3, angle=0]{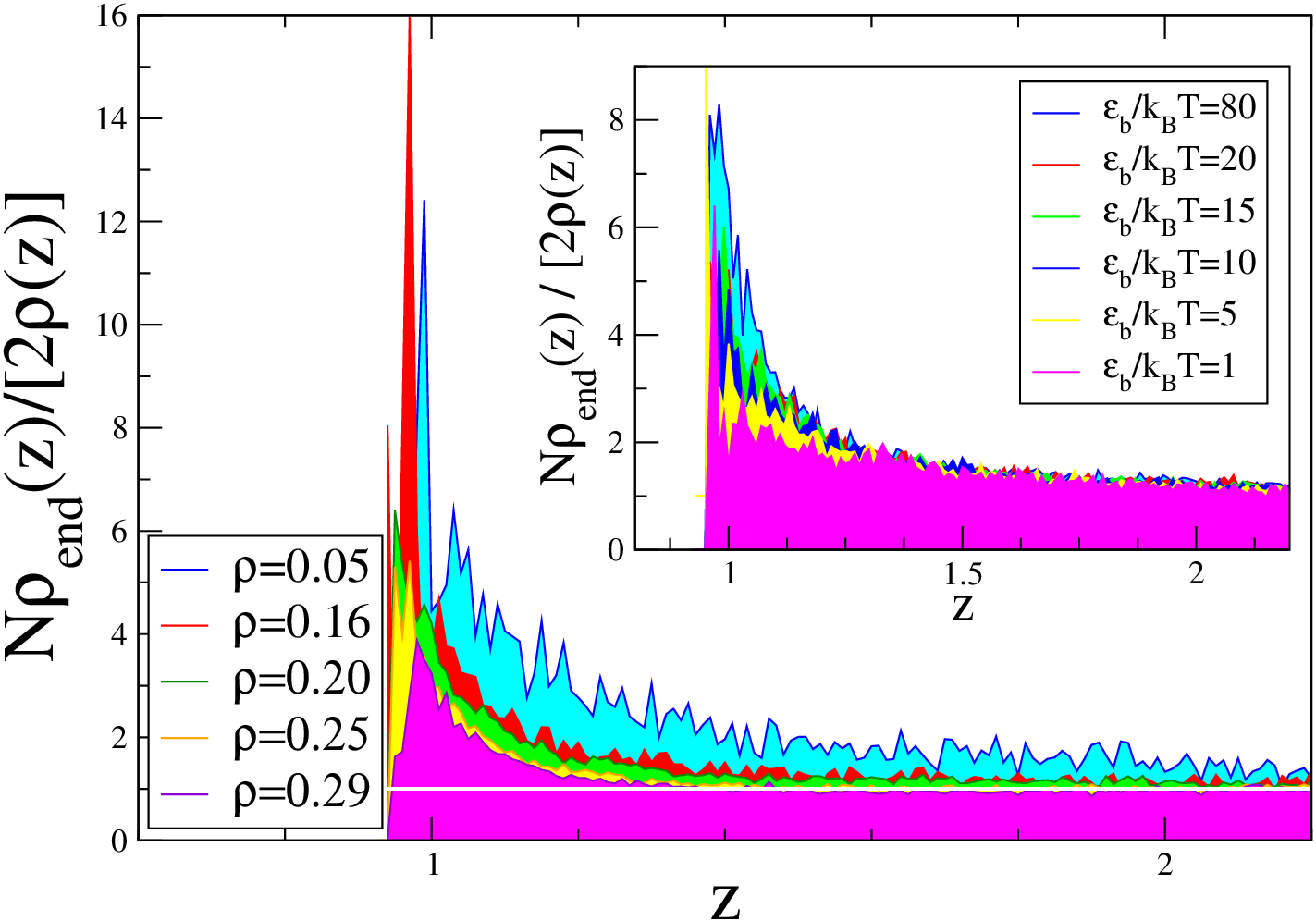}
\hspace{0.9cm}
\includegraphics[scale=0.3, angle=0]{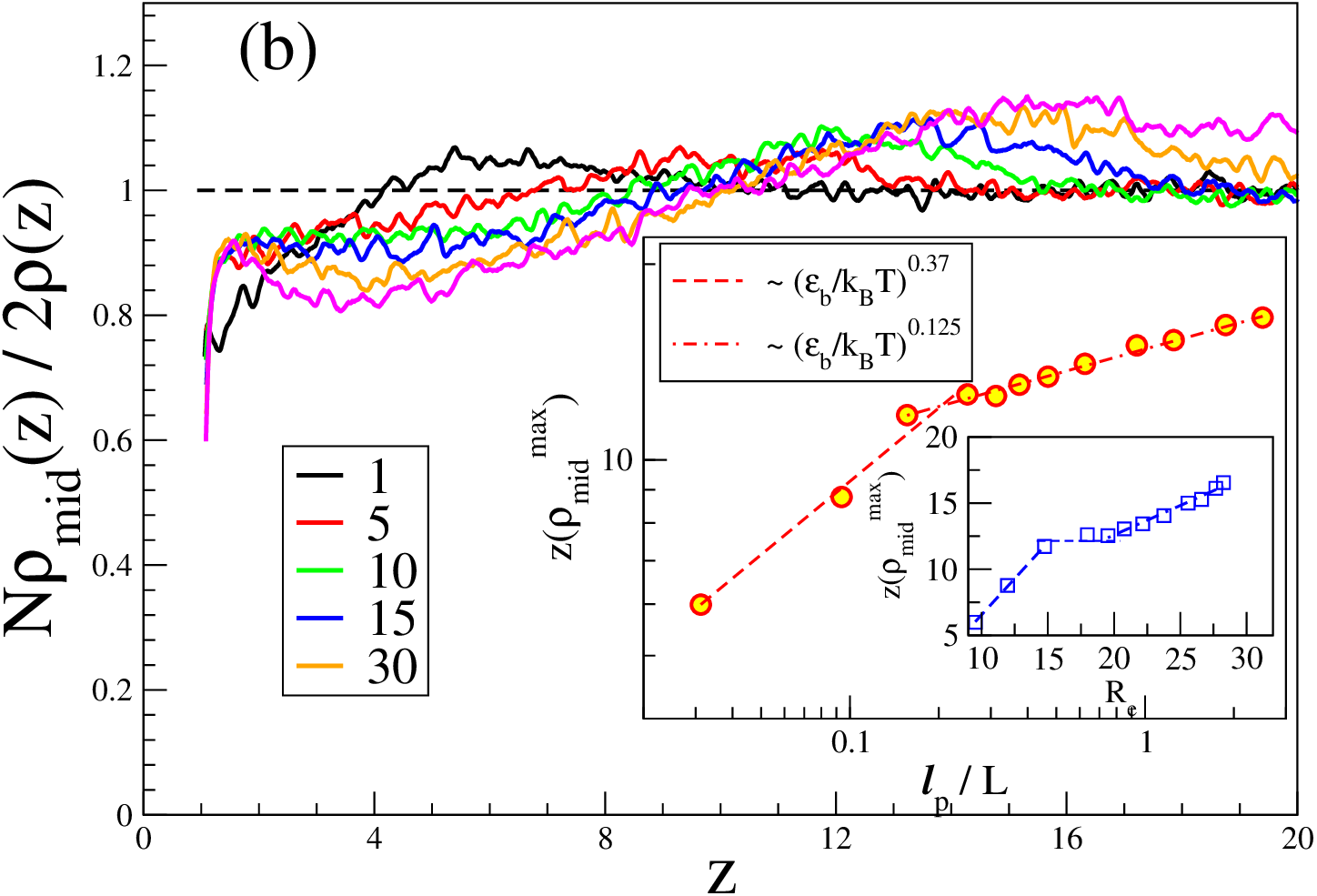}
\vspace{0.3cm}
\caption{ (a) MD results for the normalized density of end-monomers in the 
vicinity of a wall for $N=32,\; \epsilon_b/k_BT = 32$ vs $z$ for different total 
monomer concentration $\rho$ and $L_z=100$. Inset shows 
$N\rho_{end}(z)/[2\rho(z)]$ against $z$ for several values of the chain 
stiffness $\epsilon_b/k_BT$ and $L_z=40$. (b) MD results for the normalized 
density of mid-monomers for $N=$32, $\rho_b$ = 0.1 vs $z$ for several values of 
the chain stiffness $\epsilon_b/k_BT$ as indicated in the legend. Inset shows 
the location of the maximum vs $\ell_p/L$; the inset of the inset shows the 
location of the maximum vs $R_e$.} \label{fig9}
\end{figure}

\begin{figure}[htb]
\vspace{0.7cm}
\includegraphics[scale=0.26, angle=0]{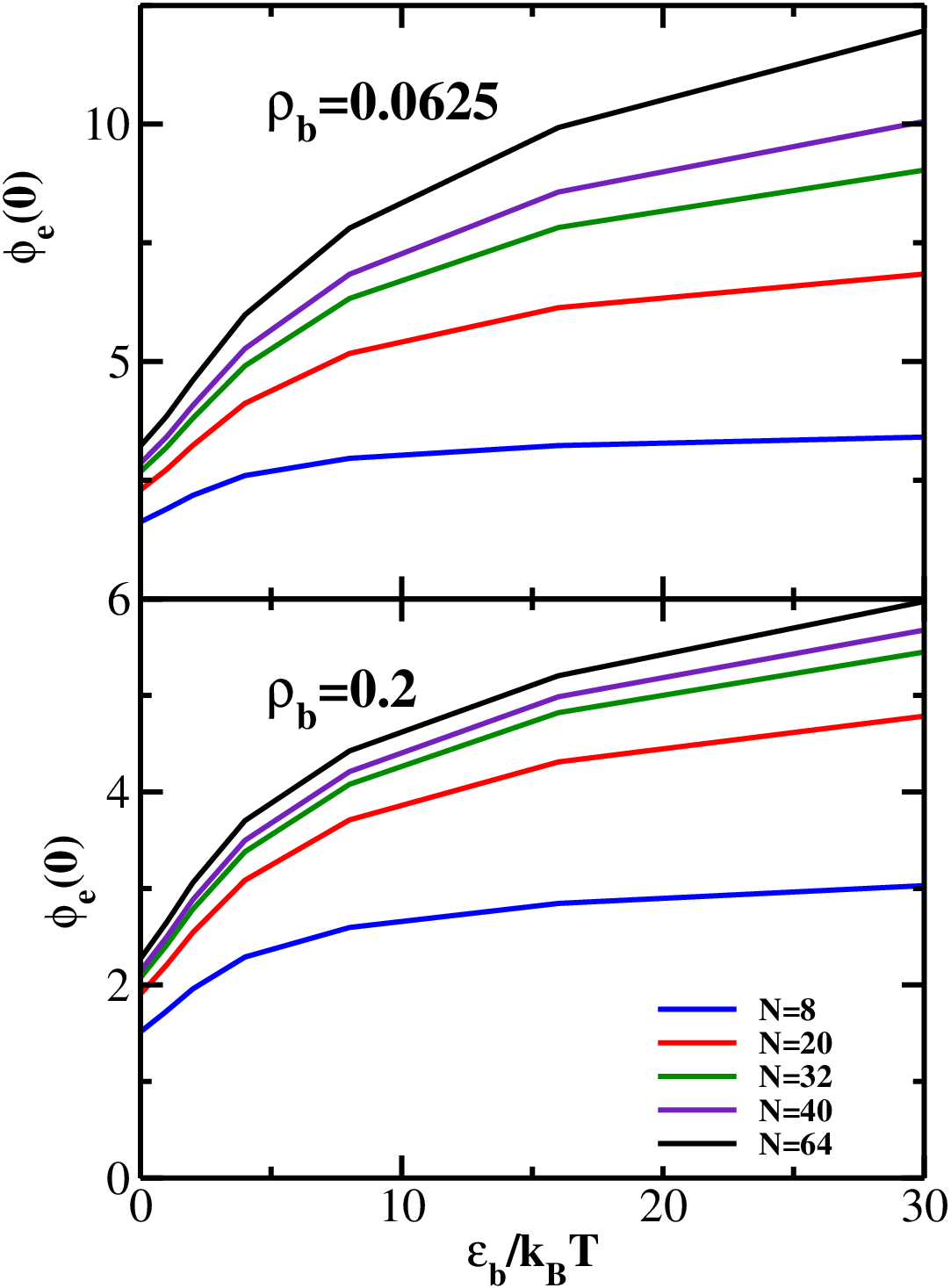}
\hspace{0.5cm}
\includegraphics[scale=0.26, angle=0]{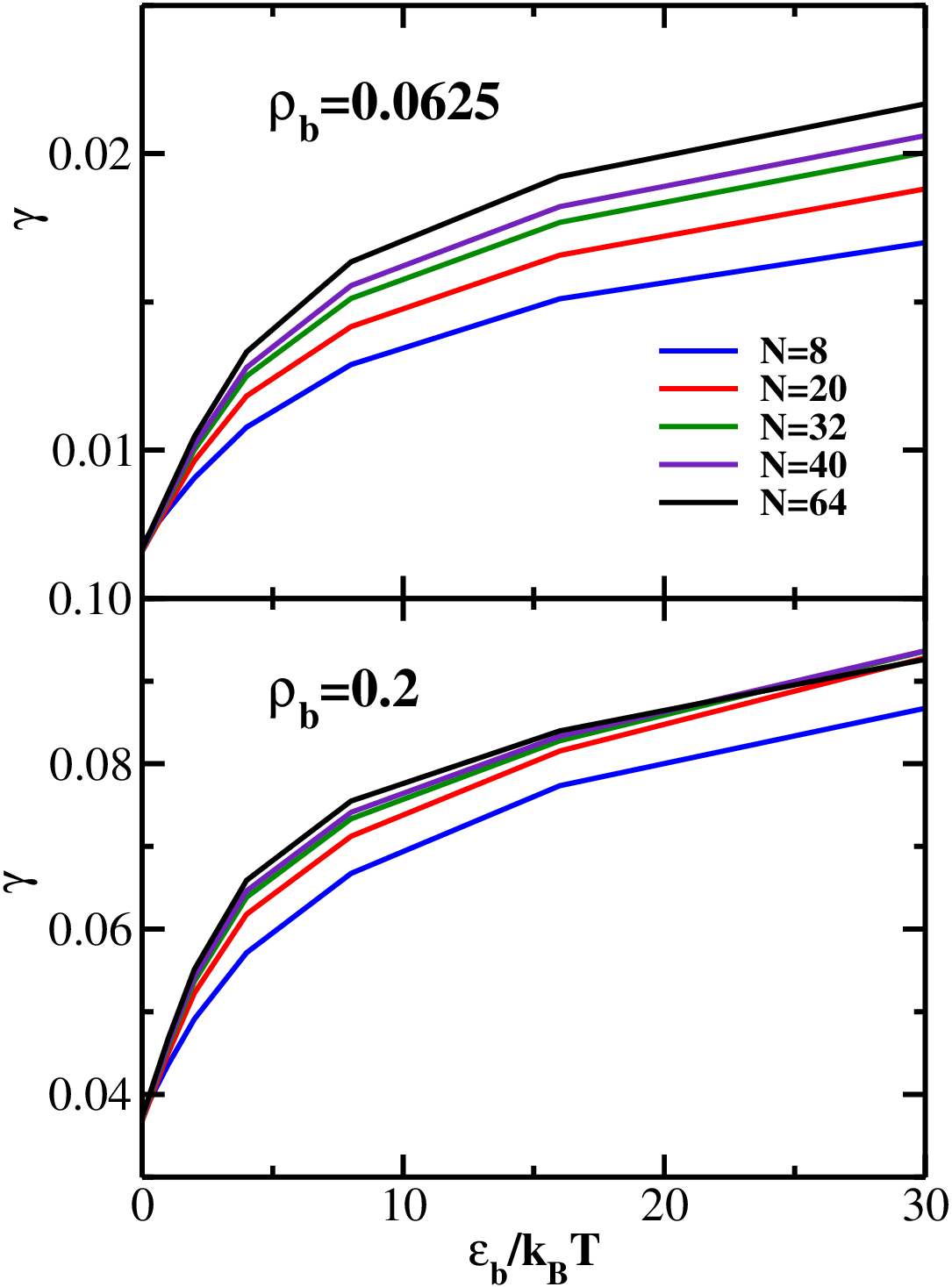}
\vspace{0.5cm}
\caption{(a) Upper panel: The DFT results for the contact
value $\phi_e(0)$ as a function of the chain stiffness  for six values of the
chain length: $N$=6, 8, 12, 16, 24, and 32; $\rho_{b}$=0.065. Lower panel: same
for $\rho_{b}$=0.2. (b) Upper panel:  The DFT results for the surface tension as
a function of the chain stiffness for six values of the chain length: $N$=6, 8,
12, 16, 24, and 32; $\rho_{b}$=0.065. Lower panel: same for $\rho_{b}$=0.2.}
\label{fig10}
\end{figure}
Given that the end-monomers are segregated to the wall and depleted away from 
the wall, one would expect the opposite to hold for the middle monomers. This is 
confirmed in the lower panel of Fig.~\ref{fig7}b, which shows the DFT results 
for the normalized middle-monomer density profiles (segments number $i$=16 and 
17 for $N$=32). There is indeed a noticeable enhancement of the middle-monomer 
density away from the wall, which increases with the chain stiffness. This 
enhancement helps to explain the weak maximum observed in $\rho(z)$ in 
Fig.~\ref{fig7}a away from the depletion zone. In order to confirm the above DFT 
predictions regarding the spatial distributions of end- and mid-monomers, we 
present in Fig.~\ref{fig9} the corresponding MD results, which all show the same 
trends as the DFT data. Of course, in MD work the price that one has to pay in 
order to have a very fine spatial resolution in $z$ are significant statistical 
fluctuations, which are absent in DFT. 

All the DFT results reported so far were limited to one particular chain length 
($N$=32) and a single value of the monomer bulk density ($\rho_{b}$=0.1). One 
major advantage of the DFT approach is its computational efficiency, which 
allows a relatively fast exploration of the parameter space (furthermore, the 
accuracy of the present DFT approach has been confirmed via comparisons with the 
corresponding MD results). In order to exploit this advantage, we present in 
Fig.~\ref{fig10}a the DFT results for the contact value $\phi_e(0)$ as a 
function of the chain stiffness for six values of the chain length: $N$=6, 8, 
12, 16, 24, and 32. The upper panel displays the results for the bulk monomer 
density $\rho_b=0.065$, while the lower panel gives the results for 
$\rho_b=0.2$. One sees that the segregation of the chain ends to the surface 
increases monotonocally with the chain stiffness for all the chain lengths 
considered. For a given value of $\epsilon_{b}/k_BT$, the segregation increases 
with  increasing chain length and with decreasing bulk density.
 
In Fig.~\ref{fig10}b, we present a similar set of the DFT results for the 
surface tension as a function of $\epsilon_{b}/k_BT$ for six values of the chain 
length and two values of the bulk monomer density. Similar to the behavior of 
$\phi_e(0)$, $\gamma$ increases monotonically with $\epsilon_{b}/k_BT$ for all 
the values of $N$ and $\rho_b$. For a given value of $\epsilon_{b}/k_BT$, the 
surface tension increases with the chain length for stiffer chains 
($\epsilon_{b}/k_BT \ge 2$), while for more flexible chains, the opposite trend 
is observed.

\subsection{The effect of varying chain length} 
\label{subsection4b}

Next, we consider the effect of varying chain length at a fixed value of the
stiffness parameter. The DFT results for the total monomer density profiles
(normalized by the bulk monomer density) are shown in the upper panel of
\begin{figure}[htb]
\vspace{0.7cm}
\includegraphics[scale=0.3, angle=0]{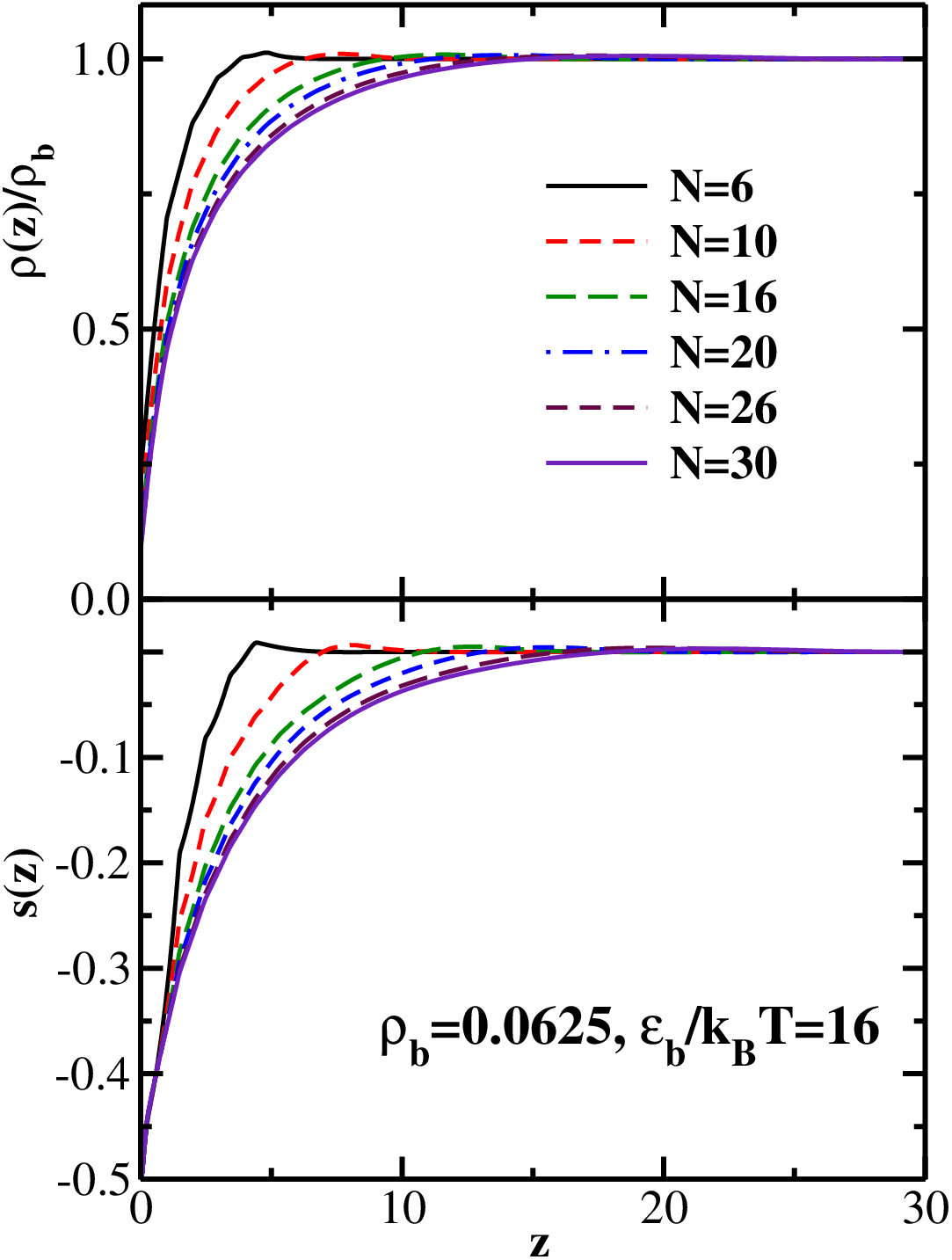}
\hspace{0.7cm}
\includegraphics[scale=0.3, angle=0]{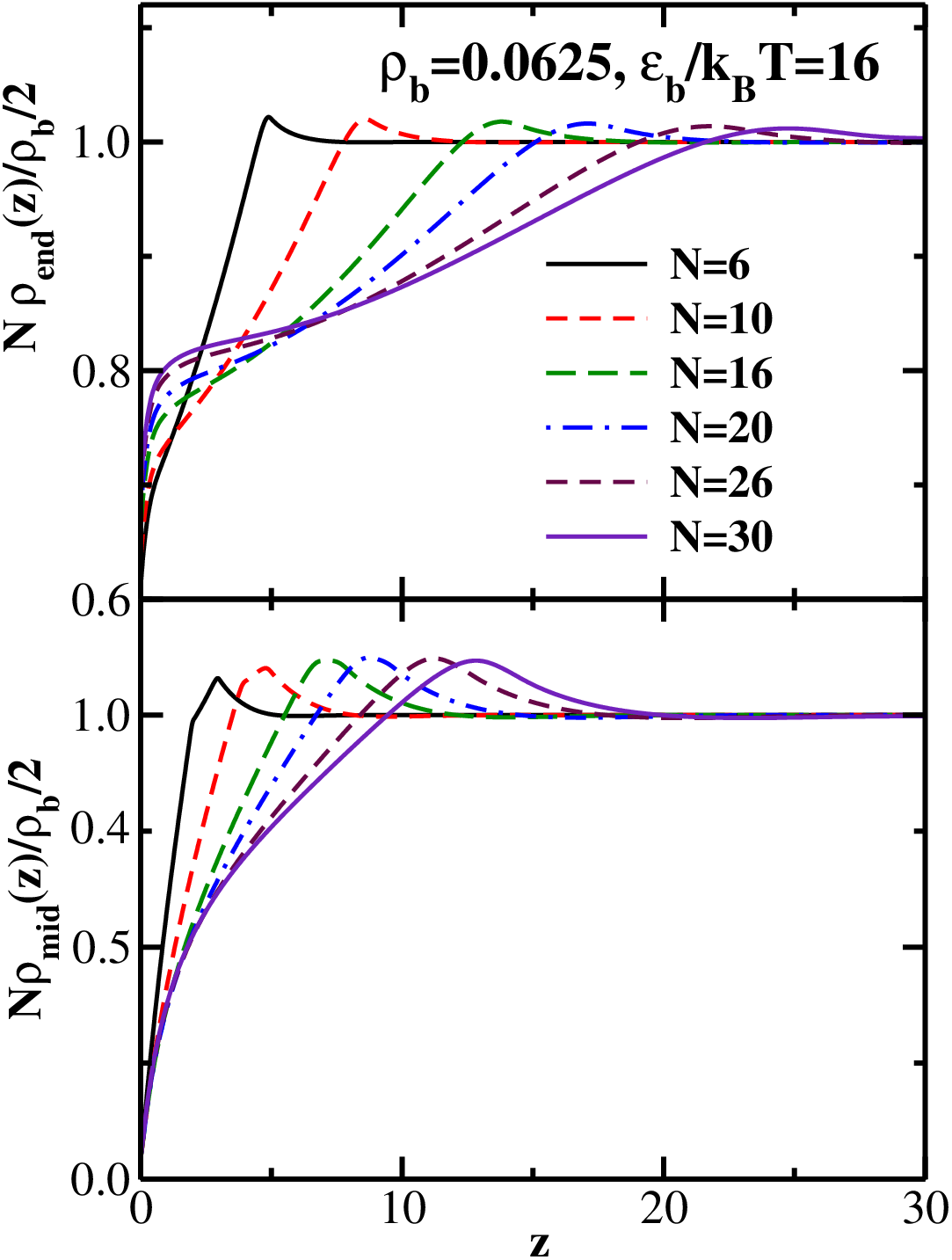}

\caption{(a) Upper panel: the normalized monomer density profiles 
$\rho(z)/\rho_{b}$ for six values of the chain length: $N$=6, 10, 16, 20, 26, 
and 30; $\epsilon_{b}/k_BT$=16 and $\rho_{b}$=0.0625. Lower panel: same for the 
bond orientational order parameter $S(z)$. (b) Upper panel: the normalized 
end-monomer density profiles for six values of the chain length: $N$=6, 10, 16, 
20, 26, and 30; $\epsilon_{b}/k_BT$=16 and $\rho_{b}$=0.0625. Lower panel: same 
for the two middle monomers.}
\label{fig11} 
\end{figure}
Fig.~\ref{fig11}a for several values of the chain length, for 
$\epsilon_{b}/k_BT$=16 and $\rho_b$=0.0625. As one would expect, the range of 
the depletion zone grows with increasing chain length, which leads to increasing 
surface tension. In the lower panel of Fig.~\ref{fig11}a we show DFT results for 
the bond orientational order parameter $S(z)$. One sees that with increasing 
chain length the tendency of the chains to be aligned parallel to the wall 
extends to larger distances from the wall.  

\begin{figure}[htb]
\vspace{0.7cm}
\includegraphics[scale=0.3, angle=0]{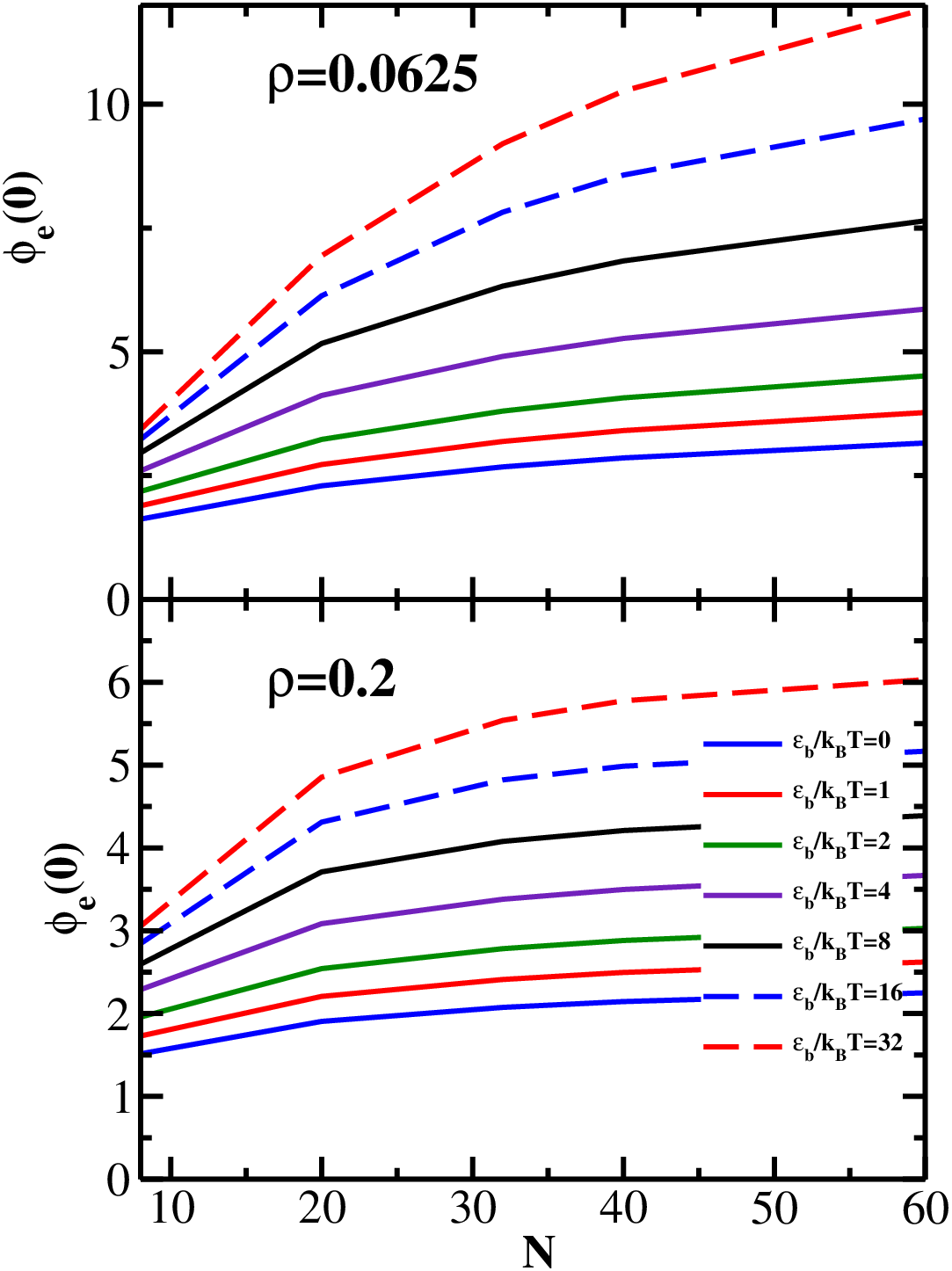}
\hspace{0.9cm}
\includegraphics[scale=0.3, angle=0]{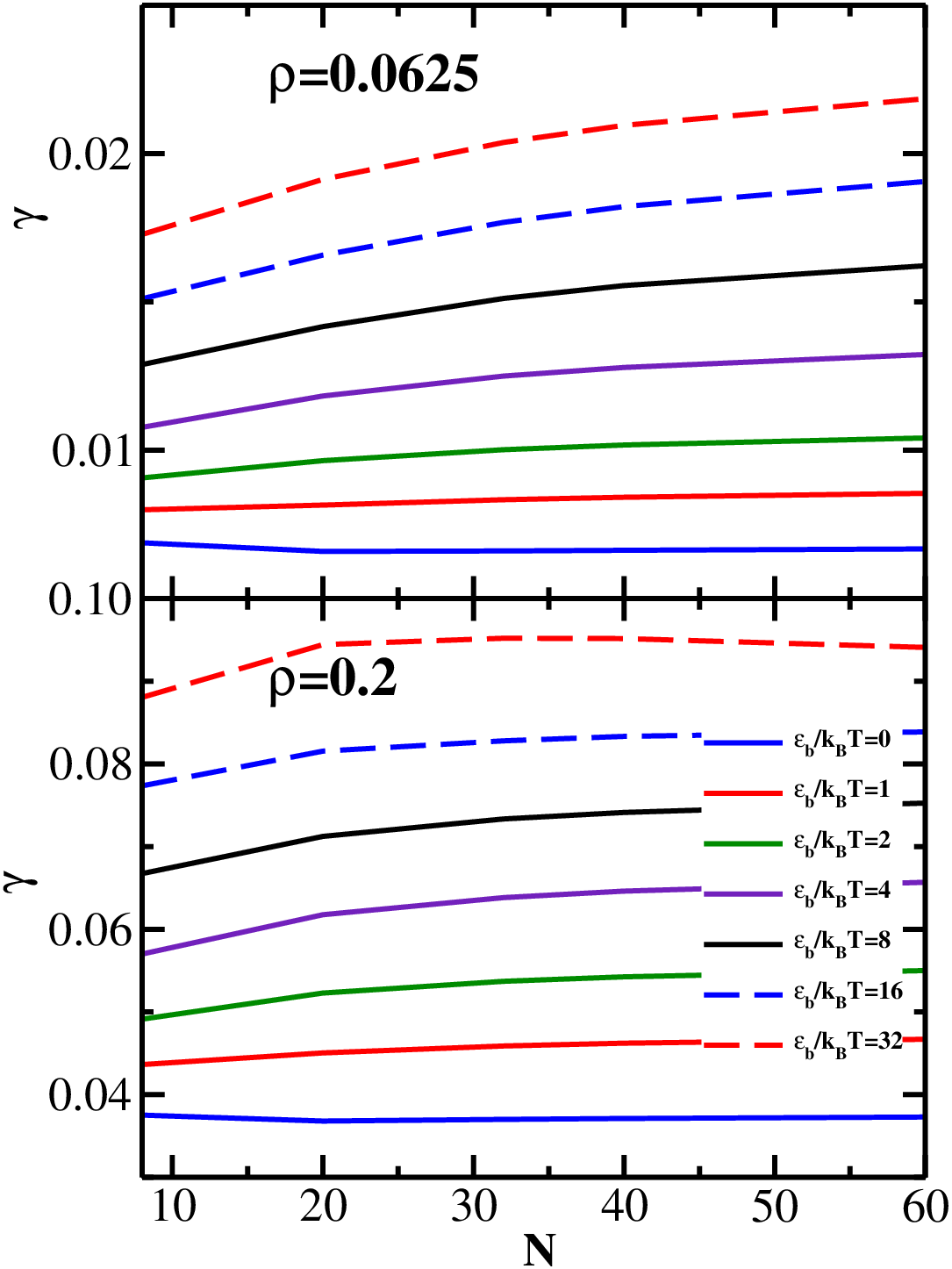}

\caption{(a) Upper panel:  The DFT results for the contact value $\phi_e(0)$ as 
a function of the chain length for nine values of the chain stiffness: 
$\epsilon_{b}/k_BT$=0, 1, 2, 3, 5, 8, 16, 24, and 32; $\rho_{b}$=0.0625. Lower 
panel: same for $\rho_{b}$=0.2. (b) Upper panel:  The DFT results for the 
surface tension as a function of the chain length for nine values of the chain 
stiffness: $\epsilon_{b}/k_BT$=0, 1, 2, 3, 5, 8, 16, 24, and 32; 
$\rho_{b}$=0.0625. Lower panel: same for $\rho_{b}$=0.2.} 
\label{fig12}
\end{figure}

Moving next to the density profiles of individual monomers, in the upper panel 
of Fig.~\ref{fig11}b we display the DFT results for the end-monomer density 
profiles $\rho_{end}(z)$ (normalized by their bulk values), while the lower 
panel shows the corresponding middle-monomer profiles. Once again, the 
segregation of the chain ends to the wall and the enhancement of the 
middle-monomer density away from the wall is quite evident. Fig.~\ref{fig8}b 
plots the ratio $\phi_e(z)$ for several values of the chain length, and one 
observes that the contact value $\phi_e(0)$ grows strongly with increasing $N$ 
(note that the contact value of the total monomer density $\rho(0)$, i.e. the 
bulk pressure, decreases with $N$).  To illustrate this behavior for other 
values of the stiffness parameter, Fig.~\ref{fig12}a displays the DFT results 
for the contact value $\phi_e(0)$ as a function of the chain length for nine 
values of the chain stiffness: $\epsilon_{b}/k_BT$=0, 1, 2, 3, 5, 8, 16, 24, and 
32; the upper panel is for the monomer bulk density $\rho_{b}$=0.0625, while the 
lower panel is for $\rho_{b}$=0.2. One sees that the segregation of the chain 
ends to the surface increases steadily with the chain length for all the values 
of $\epsilon_{b}/k_BT$  considered. For a given value of $N$, the segregation 
increases with  increasing chain stiffness and with decreasing bulk density.

In Fig.~\ref{fig12}b, we present a similar set of the DFT results for the 
surface tension as a function of $N$ for nine values of the chain stiffness and 
two values of the bulk monomer density. For the two smallest values of 
$\epsilon_{b}/k_BT$ ($\epsilon_{b}/k_BT$=0 and 1), the surface tension is seen 
to decrease with the chain length, while for stiffer chains, the opposite 
behavior is observed.

\section{Conclusions}
\label{section5}

In this paper, a comprehensive investigation of semiflexible polymers in a 
solution under good solvent conditions interacting with a repulsive flat wall 
have been presented, combining results from extensive MD simulations with a 
newly extended formulation of DFT. This new formulation was required in order to 
take into account that both the spatial density distribution is inhomogeneous 
($\rho(z)$ depends on the distance $z$ from the wall), and the angular 
distribution $f(z,\omega)$ of the orientation of the bonds is both 
spatially-dependent and anisotropic, unlike the bulk isotropic solution, where 
$\rho(z)=\rho_b$ is the given bulk density, and $f(z,\omega)=1/(4\pi)$ is also a 
constant everywhere. 

While these wall-induced inhomogeneities of $\rho(z)$ and $f(\Omega)$ mentioned
above already occur in a solution of rigid rods (at densities $\rho$ less than
the density $\rho_i$ where in the bulk the two-phase coexistence region between
isotropic (I) and nematic (N) phases begins), and it is of interest to study the
range over which these wall-induced inhomogeneities extend and to clarify their
interplay~\cite{mao97},  in a solution of semiflexible polymers additional
phenomena occur: the end-to-end vector of a chain is oriented parallel to the
wall when the center of mass of the chain is close to the wall. Related to this,
nontrivial profiles $R_{g||}^2(z),\; R_{g\bot}^2(z)$ of the parallel and
perpendicular parts of the mean-square gyration radius of the chains occur (here
$z$ means the distance of the chain center of mass from the wall). Also, chain
ends get enriched (and the density of middle monomers depleted) at the wall, if
the chain has its center of mass close to the wall. All these phenomena are
carefully quantified in our study. Also, the surface tension of the polymer
solution due to the repulsive wall is computed.

The MD simulations have utilized the standard Kremer-Grest bead-spring model, 
augmented by a bond bending potential (Eq.~\ref{eq3}). The bond bending 
potential parameter $\epsilon_b/k_BT$ (which is basically the ratio of the 
persistence length $\ell_p$  and the bond length $\ell_b$) was varied from 
$\epsilon_b/k_BT=0$ (flexible chains, where  $\ell_p\approx \ell_b$) to 
$\epsilon_b/k_BT=\ell_p/\ell_b=100$. If the chains are sufficiently stiff 
($\epsilon_b/k_BT\ge 8$), nematic order sets in at sufficiently high monomer 
concentrations (since no explicit solvent was included, $\rho$ is nothing but 
the monomer density in the simulated system). We have checked where this onset 
of the nematic order occurs (see e.g.~Fig.~2). While for short chains ($N\le 
16$) this happens only for rather concentrated solutions even if 
$\epsilon_b/k_BT$ is very large, for long chains (e.g. $N=64$) the nematic order 
sets in at rather small values of $\rho$ already in the bulk. In the present 
paper, we have deliberately avoided such densities in our simulation geometry 
(which is a slit with two equivalent repulsive wall a distance $L_z$ apart, see 
Fig.~\ref{fig1}), -- a study of capillary nematization for variable slit widths 
$L_z$ is planned for a subsequent study.

For dilute solutions, the surface tension due to the walls is very small 
(Figs.~\ref{fig3}, \ref{fig10}b, and \ref{fig12}b), but increases both with 
$\epsilon_b/k_BT$ and with chain length $N$ (in the latter case, except for very 
flexible chains). We established a reasonably good agreement between MD and DFT 
(note that the statistical accuracy of MD is a problem when the surface tension 
is very small), while the agreement for the density profiles is nearly perfect 
(Fig.~\ref{fig2}). In addition, MD confirms the DFT prediction regarding the 
enrichment of the end-monomers at the repulsive wall and its increase with 
increasing chain stiffness. All these observations give us confidence in the 
accuracy of the DFT approach.

The one-dimensional version of the DFT employed here cannot yield any 
information on the chain conformations as a whole, but this information is 
readily extracted from MD. Already in the bulk a gradual crossover from coils to 
(flexible) rods is encountered with increasing $\epsilon_b/k_BT$ 
(Fig.~\ref{fig5}). At the wall, one not only observes chain orientation parallel 
to the wall, as mentioned above (Fig.~\ref{fig6}a,b), but also the individual 
bonds of stiff chains get progressively oriented parallel to the wall, when the 
density increases (Fig.~\ref{fig6}c). This surface-induced nematic ordering 
leads to a logarithmic growth of the thickness of the surface-induced liquid 
crystalline layer at the surface as $\rho$ tends toward $\rho_i$ where in the 
bulk the ordering would set in. With increasing chain length, for stiff chains 
the thickness of the region that is affected by the wall gets influenced over a 
rather wide regime (Fig.~\ref{fig7}) even if the density is very small. This is 
reminiscent of the behavior of rigid rods near a wall (which are affected in 
their orientation over a distance equivalent to the rod length~\cite{mao97}). 

The DFT calculations corroborate these findings, showing that both the density 
and bond orientation are affected over a distance of the order of the 
persistence length, even if the density is small (Fig.~\ref{fig7}a).  
Interesting nonmonotonic density profiles for both end-monomers and 
middle-monomers  are predicted when the persistence length is rather large and 
comparable to the contour length (Fig.~\ref{fig7}b). 
While MD and DFT approaches yield qualitatively similar behavior for all the observables considered in this work, 
we note that an explicit quantitative between MD and DFT results for structural properties of the chains makes little sense since the chain models differ slightly 
(bead-spring model in MD vs tangent hard-sphere model in DFT) and also the wall potentials differ. Even for the same chain length $N$ and the same choice of 
$\epsilon_b/k_B T$, properties like $S(z)$, distributions of end-monomers or middle monomers must be slightly but systematically different. In view of the above, we present MD and DFT results side-by-side in order to demonstrate that the generic behavior is the same, irrespective of the precise choice of the model.

Finally, it is interesting to 
note that the (suitably scaled) contact value $\phi_e(0)$ of the end-monomer 
volume fraction at the wall and the surface tension have rather similar trends 
as functions of $\epsilon_b/k_BT$ and $N$ in the dilute regime 
(Figs.~\ref{fig10} and \ref{fig12}). The extended range over which stiff 
polymers ``feel'' the effect of a surface even in a dilute solution can be 
expected to have interesting consequences for the interaction of biopolymers 
(which are often rather stiff, e.g. DNA, actin etc) with biological membranes. 
Even more interesting phenomena might be expected if the entropic repulsion of 
the stiff polymers due to the wall competes with  a short-range attraction, and 
a possible adsorption transition of the semiflexible polymers 
occurs~\cite{forsman06,turesson07,birshtein79,hsu13}. We hope to address such 
issues in our future work. 

\section{Acknowledgements}
 
S.A.E. acknowledges financial support from the Alexander von Humboldt
Foundation. A.M. thanks for partial support under the grant No $BI314/24$.
Parts of this research were conducted using the supercomputer Mogon and/or advisory services offered by Johannes Gutenberg University Mainz (www.hpc.uni-mainz.de), which is a member of the AHRP and the Gauss Alliance e.V.
The authors gratefully acknowledge the computing time granted on the supercomputer Mogon at Johannes Gutenberg University Mainz (www.hpc.uni-mainz.de).

 
\end{document}